\documentclass[12pt]{article}
\usepackage{putex}
\usepackage{graphicx}
\def\half{{{1 \over 2}}}
\def\im{\operatorname{Im}}
\def\re{\operatorname{Re}}
\begin{document}

\preprint{hep-th/0611005 \\ PUPT-2210}

\institution{PU}{Joseph Henry Laboratories, Princeton University, Princeton, NJ 08544}

\title{Expanding plasmas and quasinormal modes of anti-de Sitter black holes}

\authors{Joshua J. Friess, Steven S. Gubser, Georgios Michalogiorgakis, and \\[10pt] Silviu S. Pufu}

\abstract{We compute the gravitational quasinormal modes of the global $AdS_5$-Schwarzschild solution.  We show how to use the holographic dual of these modes to describe a thermal plasma of finite extent expanding in a slightly anisotropic fashion.  We compare these flows with the behavior of quark-gluon plasmas produced in relativistic heavy ion collisions by estimating the elliptic flow coefficient and the thermalization time.}


\maketitle
\tableofcontents

\section{Introduction}
\label{INTRODUCTION}

When gold nuclei are collided at $\sqrt{s_{NN}} = 200\,{\rm GeV}$ at the Relativistic Heavy Ion Collider (RHIC), it is plausible that a thermal state forms in which quarks and gluons are deconfined.  The peak temperature is estimated to be roughly $300\,{\rm MeV}$, somewhat above the temperature $T_c$ at which QCD deconfines.\footnote{Lattice estimates for $T_c$ tend to fall in the range $160-190\,{\rm MeV}$ \cite{Muller:2006ee}.  Recent studies \cite{Cheng:2006qk,Aoki:2006br} suggest that there is still some disagreement within the lattice community about where in this range $T_c$ really lies.}  The resulting ``quark-gluon plasma'' (QGP) is thought to be strongly coupled, partly because its collective motions are well described by hydrodynamics with shear viscosity much less than the entropy density: $\eta/s \ll 1$.  Weakly coupled plasmas generally have $\eta/s \gg 1$.  Intriguingly, certain black holes in string theory which describe thermal states of strongly coupled gauge theories have $\eta/s = 1/4\pi$.  Hard probes of these black holes, in the form of classical strings trailing across their horizons, exhibit energy loss which has similarities, both in magnitude and angular distribution, to experimental results on jet-quenching.  The hope naturally arises that further and closer analogies might exist between real-world QGP's and string theoretic black holes.  Can we ``simulate'' a heavy-ion collision, from the moment of impact to the onset of hadronization, in string theory?  The aim of this paper is to take a step in this direction by describing expanding, cooling plasmas of finite extent in string theory.

Recent accounts of the discoveries at RHIC include the authoritative summaries \cite{Arsene:2004fa,Adcox:2004mh,Back:2004je,Adams:2005dq} and the recent review \cite{Muller:2006ee}.  The shear viscosity of black holes and its possible relevance to the QGP were developed in several papers including \cite{Kovtun:2004de,Policastro:2001yc}.  The theoretical framework for relating black hole physics to gauge theory is the anti-de Sitter / conformal field theory correspondence (AdS/CFT), formulated in \cite{Maldacena:1997re,Gubser:1998bc,Witten:1998qj} and reviewed (among other places) in \cite{MAGOO,DHoker:2002aw}.  One of the first hints of the gauge-string duality was the calculation of the entropy density of D3-branes \cite{Gubser:1996de}, which feeds into the calculations of shear viscosity to entropy density and also shows a $25\%$ pressure deficit that is reminiscent of lattice results for QCD (see for example \cite{Karsch:2001cy}).

The literature on the possible connection between string theory and RHIC physics has grown large.  The possibility of emulating a heavy ion collision in an AdS/CFT context was emphasized already in \cite{Nastase:2005rp,Shuryak:2005ia,Nastase:2006eb}, and the connection with finite-sized black holes is already present in these papers.  Bjorken flow in an infinite medium has been studied from an AdS/CFT perspective, for example in \cite{Janik:2005zt,Nakamura:2006ih}; see also \cite{Sin:2006pv}.  Quasinormal modes have been considered in \cite{Kovtun:2005ev,Kovtun:2006pf,Teaney:2006nc,Janik:2006gp}, the first of which especially has results that are analogous in significant respects to ours, albeit in an infinite static background.  Hard probes were studied in \cite{Liu:2006ug,Herzog:2006gh,Casalderrey-Solana:2006rq,Gubser:2006bz,Friess:2006aw,Friess:2006fk} and compared with some success to the phenomenon of jet-splitting \cite{Adler:2005ee,Adams:2005ph}.  Quarkonium systems have also been studied: see for example \cite{Herzog:2006gh,Peeters:2006iu,Liu:2006nn,Chernicoff:2006hi,Argyres:2006vs}.

It is important to bear in mind that most studies of the gauge-string-RHIC connection have been carried out using ${\cal N}=4$ super-Yang-Mills theory at strong coupling.  This theory is different in many respects from QCD: in particular, it does not confine, it does not include quarks in the fundamental representation of the color group, and its vacuum structure is completely different.  The current study is particularly closely tied to the conformal invariance that ${\cal N}=4$ super-Yang-Mills possesses.  Conformal invariance implies an equation of state $p = \epsilon/3$, so the speed of sound is $v=1/\sqrt{3}$.  The average speed of sound in the QGP produced at RHIC is believed to be substantially lower: for example, an average value $v=1/3$ is used in \cite{Casalderrey-Solana:2004qm}.  The assumption of conformal invariance assumption may be better justified at the highest temperatures attained in a RHIC collision, and it is possible that a broader conformal regime will arise in LHC heavy-ion collisions.

The point of entry for the present paper is the observation that the holographic image on Minkowski space of the global $AdS_5$-Schwarzschild black hole is a spherical shell of plasma first contracting and then expanding.\footnote{The dual description of global $AdS_5$-Schwarzschild has been considered previously in \cite{Horowitz:1999gf} using conformal transformation methods similar to those described in section~\ref{RADIAL}.  We thank G.~Horowitz for bringing \cite{Horowitz:1999gf} to our attention.}  We describe this gauge theory configuration as a ``conformal soliton flow'' because the expanding shell has exactly the same profile as the contracting one, as if there were no interactions.  A flow with the same stress-energy tensor could be arranged in a free massless theory.  It is appropriate to describe the soliton as conformal because its existence relies almost entirely on the ability to make a conformal transformation from part of $S^3 \times {\bf R}$ to ${\bf R}^{3,1}$.  In $S^3 \times {\bf R}$, the conformal soliton is simply a plasma at rest.

Having established this spherical ``approximation'' to a heavy-ion collision in section~\ref{RADIAL}, we proceed to study its small deformations in linear perturbation theory.  These deformations are known as quasinormal modes (QNM's).  A large literature is devoted to them, stretching back to the classic computations \cite{Regge57,Zerilli:1970se} for the four-dimensional Schwarzschild black hole.  The method of choice for numerically determining quasinormal modes is usually to factor out the asymptotic near-horizon behavior and expand the rest in a power series around the horizon.  Then a convergence property in the far-field limit determines $\omega$.  Subtle methods were developed in \cite{1993PhRvD..47.5253N}, following earlier work \cite{Leaver:1985ax}, to handle the convergence issues in a numerically robust manner.  A more straightforward approach, which is often sufficient for low-lying normal modes, was first used in the GAdSBH background in \cite{Horowitz:1999jd} and in \cite{Cardoso:2003cj} to treat gravitational perturbations of global $AdS_4$-Schwarzschild.  Asymptotic expressions based on a different method (related to that of \cite{Motl:2003cd}) were obtained in \cite{Siopsis:2004up} for global $AdS_5$-Schwarzschild, and in \cite{Musiri:2005ev} for large black holes in anti-de Sitter space; but we are unaware of a full treatment of the case of interest, namely gravitational perturbations of the global $AdS_5$-Schwarzschild black hole (GAdSBH).

The plan of sections~\ref{TENSOR}, \ref{VECTOR}, and~\ref{SCALAR} is to examine in detail the tensor, vector, and scalar quasinormal modes of the GAdSBH and extract from them the dual perturbations to the conformal soliton flow \eno{FoundBackgroundT}.  In principle, this constitutes a complete analysis of small perturbations.  Special interest attaches to the low-lying QNM's because they tend to dominate late-time behavior.

Building blocks for the quantitative study of QNM's are summarized in section~\ref{BUILDING}.  Sections~\ref{FITS} and~\ref{SUMMARY} summarize the main calculations.  In section~\ref{CASES} we consider some explicit examples of deformed conformal soliton flows that have some features in common with heavy-ion collisions.  In section~\ref{DISCUSS} we discuss the linearized hydrodynamic approximation to the low-lying quasinormal modes as well as the motion of test masses in $AdS_5$.  We also speculate in section~\ref{DISCUSS} about colliding black holes in $AdS_5$.  An appendix is devoted to an exposition of scalar, vector, and tensor spherical harmonics on $S^3$.

The reader wishing to gain a first impression of our results without reading through all the details may wish to read sections~\ref{BACKGROUND}, \ref{RADIAL}, \ref{SIZE}, \ref{SUMMARY}, \ref{SLOWSCALAR}, \ref{FASTMODES}, and~\ref{CONCLUSIONS}.  Our estimates of the elliptic flow coefficient and the thermalization time can be found in sections~\ref{SLOWSCALAR} and~\ref{FASTMODES}, respectively, and section~\ref{CONCLUSIONS} is an executive summary of our results.

\section{Building blocks of the calculation}
\label{BUILDING}

\subsection{The background metric}
\label{BACKGROUND}

The global $AdS_5$-Schwarzschild black hole, hereafter abbreviated as GAdSBH, takes the form
 \eqn{GAdSBHmetric}{
  ds^2 = -\left( 1 - {\rho_0^2 \over \rho^2} +
    {\rho^2 \over L^2} \right) d\tau^2 +
   {d\rho^2 \over 1 - {\rho_0^2 \over \rho^2} +
    {\rho^2 \over L^2}} + \rho^2 d\hat\Omega^2 \,,
 }
where $d\hat\Omega^2$ is the line element on a unit $S^3$:
 \eqn{sThreeMet}{
  d\hat\Omega^2 = \hat{g}_{ij} dy^i dy^j =
    d\chi^2 + \sin^2 \chi (d\theta^2 + \sin^2 \theta \, d\phi^2) \,.
 }
The background \eno{GAdSBHmetric} is a solution to the vacuum Einstein equations augmented by a cosmological term:
 \eqn{VacuumEinstein}{
  R_{ab} + {4 \over L^2} g_{ab} = 0 \,.
 }
These equations of motion follow from the action
 \eqn{FiveDaction}{
  S = {1 \over 2\kappa^2} \int d^5 x \, \sqrt{g} \left[ R +
    {12 \over L^2} \right] \,.
 }
This action captures part of the dynamics of type~IIB string theory compactified on an $S^5$ of radius $L$ with $N$ unit of Ramond-Ramond five-form flux, where
 \eqn{LkappaRelation}{
  L^3/\kappa^2 = (N/2\pi)^2 \,.
 }
AdS/CFT relates the GAdSBH background to a thermal state of $SU(N)$ ${\cal N}=4$ super-Yang-Mills (SYM) on $S^3 \times {\bf R}$.\footnote{In this simplest example of AdS/CFT, the five-dimensional gravitational constant is expressed in terms of the ten-dimensional one as $\kappa^2 = \kappa_{10}^2/\pi^3 L^5$.  In more complicated examples, both the \eno{LkappaRelation} and the relation between $\kappa^2$ and $\kappa_{10}^2$ change.}

The location $\rho_H$ of the horizon of GAdSBH is the most positive root of $f(\rho) = 0$.  It is related to $\rho_0$ by the following two equivalent expressions:
 \eqn{rhoHdef}{
  \rho_H = L \sqrt{\sqrt{1 + 4 \rho_0^2/L^2} - 1 \over 2} \qquad
   \rho_0 = \rho_H \sqrt{1 + \rho_H^2/L^2} \,.
 }
For future reference, let us record the standard results for the mass, entropy, and temperature:
 \eqn{GAdSBHmts}{
  M = {3\pi^2 \over \kappa^2} \rho_0^2 \qquad
  S = {4\pi^3 \rho_H^3 \over \kappa^2} \qquad
  T = {\rho_H \over \pi L^2} \left( 1 + {L^2 \over 2\rho_H^2}
    \right) \,.
 }
The mass is interpreted as the total energy in the boundary gauge theory (modulo a tem\-per\-a\-ture-independent Casimir contribution), and it may be calculated explicitly by integrating the energy density $\langle \tilde{T}^{\rm Sch}_{00} \rangle$ of \eno{FoundSthreeT} over the $S^3$ of radius $L$ on which the boundary gauge theory resides.

We are interested in perturbations of \eno{GAdSBHmetric} which solve the linearized Einstein equations following from \eno{VacuumEinstein}.  The perturbations should be infalling at the horizon, and they should have the fall-off near the boundary of GAdSBH corresponding to an expectation value of the gauge theory stress tensor $T_{mn}$ rather than an alteration of the metric (see section~\ref{HOLOGRAPHIC} for more detail).  If a time-dependence of the form $e^{-i\omega \tau}$ is assumed, then $\Im\omega < 0$ corresponds to stable modes.\footnote{Some works on QNM's assume a time-dependence $e^{i\omega t}$ and find stability when $\Im\omega > 0$.  The true field configurations are of course made real by combining complex amplitudes with their conjugates.}  It is generally understood that QNM's dominate the approach to equilibrium of perturbed black holes.

\subsection{Spherical harmonics}
\label{SPHERICAL}

The strategy for solving the linearized Einstein equations in the GAdSBH background \eno{GAdSBHmetric} will be separation of variables.  So one requires generalizations of the usual spherical harmonics $Y_{\ell m}(\theta,\phi)$ appropriate to scalars, vectors, and symmetric tensors on $S^3$.  These harmonics are defined by the equations
 \eqn{DefineHSHscalar}{
  (\hat\nabla_i \partial^i + k_S^2) \mathbb{S} = 0
 }
 \eqn{DefineHSHvector}{
  (\hat\nabla_i \hat\nabla^i + k_V^2) \mathbb{V}_j = 0 \qquad
    \hat\nabla^j \mathbb{V}_j = 0
 }
 \eqn{DefineHSHtensor}{
  (\hat\nabla_i \hat\nabla^i + k_T^2) \mathbb{T}_{jh} &= 0 \qquad
    \hat\nabla^j \mathbb{T}_{jh} = 0 = \mathbb{T}^j{}_j \,,
 }
where $\hat\nabla$ is the standard connection on the unit $S^3$, indices are raised and lowered using the metric on the unit $S^3$.  A complete and explicit listing of the relevant harmonics is given in Appendix~\ref{TOMITA}, following \cite{Tomita:1982ew}.

\subsection{The linearized Einstein equations}
\label{LINEARIZED}

Decoupled forms of the linearized Einstein equations were derived in \cite{Kodama:2003jz}, following earlier work (see for example \cite{Kodama:2000fa}).  We will now briefly review the results, specialized to five dimensions and spherical symmetry.  First, split coordinates $y^A = (\tau,\rho,\chi,\theta,\phi)$ into $y^\alpha = (\tau,\rho)$ and $y^i = (\chi,\theta,\phi)$.  Let $D_\alpha$ denote the covariant derivative corresponding to the metric for the ``orbit space,''
 \eqn{twoDmetric}{
  ds_2^2 \equiv -f d\tau^2 +
    {1 \over f} d\rho^2 \qquad
   f = 1 - {\rho_0^2 \over \rho^2} +
     {\rho^2 \over L^2} \,.
 }
This metric is asymptotically $AdS_2$.  The overall strategy is to express graviton perturbations in terms of spherical harmonics and ``master fields,'' denoted $\Phi$, which are scalars in the orbit space \eno{twoDmetric}.  Let $\epsilon_{\alpha\beta}$ be the anti-symmetric tensor for the metric \eno{twoDmetric}, such that $\epsilon_{\tau\rho} = 1$.  For tensor modes,
 \eqn{TensorPerturb}{
  \delta g_{\alpha\beta} = 0 = \delta g_{\alpha i} \qquad
   \delta g_{ij} = 2\rho^2 \, H_T(\tau,\rho) \,
   \mathbb{T}_{ij}(\chi,\theta,\phi)
 }
 \eqn{TensorPhi}{
  H_T = \rho^{-3/2} \Phi
 }
 \eqn{TensorEOM}{
  \left( D_\alpha \partial^\alpha -
    {V_T(\rho) \over f} \right) \Phi = 0
 }
 \eqn{TensorPotential}{
  V_T(\rho) = {f \over \rho^2} \left(
    {15 \over 4} f + {6 \rho_0^2 \over \rho^2} + k_T^2 - 1
   \right) \,.
 }
For vector modes,
 \eqn[c]{VectorPerturb}{
  \delta g_{\alpha\beta} = 0 \qquad
   \delta g_{\alpha i} = \rho \, f_\alpha(\tau,\rho) \,
     \mathbb{V}_i(\chi,\theta,\phi)  \cr
   \delta g_{ij} = -{2 \over k_V} \rho^2 \, H_T(\tau,\rho) \,
    \hat\nabla_{(i} \mathbb{V}_{j)}(\chi,\theta,\phi)
 }
 \eqn{VectorPhi}{
  F_\alpha \equiv f_\alpha + {\rho \over k_V} \partial_\alpha H_T
   = {1 \over \rho^2} \epsilon_{\alpha\beta} \partial^\beta (\rho^{3/2} \Phi)
 }
 \eqn{VectorEOM}{
  \left( D_\alpha \partial^\alpha -
    {V_V(\rho) \over f} \right) \Phi = 0
 }
 \eqn{VectorPotential}{
  V_V(\rho) = {f \over \rho^2} \left(
    {3 \over 4} f - {6 \rho_0^2 \over \rho^2} + k_V^2 + 1
   \right) \,.
 }
The quantities $H_T$ in \eno{VectorPerturb} and $\Phi$ in \eno{VectorPhi} have nothing to do with the $H_T$ in \eno{TensorPerturb} and $\Phi$ in \eno{TensorPhi}.  For scalar modes,
 \eqn[c]{ScalarPerturb}{
  \delta g_{\alpha\beta} = f_{\alpha\beta} \,
    \mathbb{S}(\chi,\theta,\phi) \qquad
   \delta g_{\alpha i} = \rho f_\alpha \,
     \mathbb{S}_i(\chi,\theta,\phi)  \cr
   \delta g_{ij} = 2 \rho^2 \left[ H_L(\tau,\rho) \, \hat{g}_{ij} \,
    \mathbb{S}(\chi,\theta,\phi) + H_T(\tau,\rho) \,
    \mathbb{S}_{ij}(\chi,\theta,\phi) \right]
 }
 \eqn[c]{ScalarSDefs}{
  \mathbb{S}_i = -{1 \over k_S} \partial_i \mathbb{S} \qquad
   \mathbb{S}_{ij} = {1\over k_S^2} \hat\nabla_i \partial_j \mathbb{S} + {1\over 3} \hat g_{ij} \mathbb{S}
 }
 \eqn[c]{ScalarParams}{
  H = m + 6 w \qquad w = {\rho_0^2 \over \rho^2 } \qquad m = k_S^2 - 3
 }
 \eqn[c]{ScalarPhiDefs}{
   X_\alpha = {\rho \over k_S}\left(f_\alpha + {\rho \over k_S} \partial_\alpha H_T\right) \cr
   F_{\alpha \beta} = f_{\alpha \beta} + D_\alpha X_\beta + D_\beta X_\alpha \cr
   F = H_L + {1 \over 3} H_T + {1 \over \rho} \left(\partial^\alpha \rho \right) X_\alpha
 }
 \eqn{ScalarPhi}{
   \mathcal{F}_{\alpha \beta} \equiv F_{\alpha \beta} + g_{\alpha \beta} F = {1 \over \rho H} \left( D_\alpha \partial_\beta \left(\rho^{3/2} H \Phi\right) - \half g_{\alpha \beta} D_\gamma \partial^\gamma \left(\rho^{3/2} H \Phi\right)\right)
 }
 \eqn{ScalarEOM}{
  \left( D_\alpha \partial^\alpha -
    {V_S(\rho) \over f} \right) \Phi = 0
 }
 \eqn{ScalarPotential}{
  V_S(\rho) &= { f \over 4 \rho^2 H^2} \Big(2 \big[ 2 m^3 + 36 (7 - 3w)w^2 + m^2(7+5w) \cr &\qquad{}+12 m w (11w-9)\big] + (m^2 + 108 m w - 540 w^2)\left(2 - 2w - f \right)\Big) \,.
 }
Again, the quantities $H_T$ in \eno{ScalarPerturb} and $\Phi$ in \eno{ScalarPhi} have nothing to do with the quantities of the same names entering into the description of tensor and vector modes.

The quasinormal mode calculation consists of solving the master equations \eno{TensorEOM}, \eno{VectorEOM}, and~\eno{ScalarEOM}.  These calculations are the topics of sections~\ref{TENSOR}-\ref{SCALAR}.  To convert these calculations into predictions for the boundary gauge theory, we need expressions for the holographic stress tensor in terms of the asymptotics of the master field near the boundary of GAdSBH.  Suitable expressions are developed in sections~\ref{HOLOGRAPHIC}, \ref{TENSORSTRESS}, \ref{VECTORSTRESS}, and \ref{SCALARSTRESS}, using equations~\eno{TensorPerturb}-\eno{ScalarPotential}

\subsection{The holographic stress tensor}
\label{HOLOGRAPHIC}

Recall now the method for extracting the boundary theory stress
tensor from an asymptotically $AdS_5$ metric.  The key ingredient is the Brown-York quasi-local
stress tensor, which is defined with respect to a hypersurface
$\Sigma$ with outward normal vector field $n_A$
\cite{Balasubramanian:1999re}\footnote{The sign conventions are such
that Einstein's equations would take the form $G_{AB}=8 \pi G
T_{AB}$ if they applied (which they don't since gravity is
non-dynamical in the boundary theory).  Thus, $G_{AB}$ differs by a
sign from the one in \cite{Balasubramanian:1999re}.}:
 \eqn[c]{BYtensor}{
  T^\Sigma_{AB} = K_{AB} - K h_{AB} - {3 \over L} h_{AB} +
    {L \over 2} G_{AB}  \cr
  h_{AB} = g_{AB} - n_A n_B \qquad
  K_{AB} = -h_A{}^C \nabla_C n_B \,.
 }
Here $G_{AB}$ is the Einstein tensor of the induced metric $h_{AB}$
on $\Sigma$.  To specify $\Sigma$, one chooses a smooth, positive
function $q(y)$ of the global coordinates $y^A$ which has a simple
zero at the boundary.  Then let $\Sigma(\epsilon)$ be the locus of
points where $q(y)=\epsilon$.  The outward pointing normal form $n_A
dy^A$ is a negative multiple of $dq$.  The metric $g_{ab}$ on the
boundary and the expectation value of the stress tensor $\langle
T_{ab} \rangle$ in the boundary theory are given by
 \eqn{BMetAndT}{
  g_{ab} = \lim_{\epsilon \to 0} q^2 h_{ab} \qquad
   \langle T_{ab} \rangle = {1 \over \kappa^2}
    \lim_{\epsilon \to 0} {1 \over q^2} T_{ab}^{\Sigma(\epsilon)} \,,
 }
where $a$ and $b$ are four-valued indices along the boundary or $\Sigma(\epsilon)$ and $\kappa^2 = 8\pi G$, the five-dimensional gravitational constant appearing in \eno{FiveDaction}.

By choosing
 \eqn{zFoliator}{
  q(y) = {1 \over \sqrt{1+\rho^2/L^2}
    \cos {\tau \over L} + {\rho \over L} \cos\chi}
 }
one arranges for $g_{ab}$ to be the metric for Minkowski space, although represented in the Penrose coordinates $y^a = (\tau,\chi,\theta,\phi)$.  (Section~\ref{RADIAL} includes an explanation of the origin of \eno{zFoliator}.)  To transform to standard polar coordinates $x^m = (t,r,\theta,\phi)$, one uses
 \eqn{YtoX}{
  {t \over L} = {\sin {\tau \over L} \over
    \cos {\tau \over L} + \cos\chi} \qquad
  {r \over L} = {\sin \chi \over \cos {\tau \over L} + \cos\chi} \,.
 }
The stress tensor transforms simply when passing from $y^a$ coordinates to $x^m$ coordinates:
 \eqn{FinalT}{
  \langle T_{mn} \rangle =
    {\partial y^a \over \partial x^m}
    {\partial y^b \over \partial x^n} \langle T_{ab} \rangle \,.
 }
$\langle T_{mn} \rangle$ is the ``answer'' that we want to calculate starting from a perturbed GAdSBH background: for instance, $\langle T_{00} \rangle$ is the energy density in the boundary gauge theory, and other components relate to the pressure, the energy flow, and so forth.

In practice, it is onerous to work through \eno{BYtensor}-\eno{FinalT} for perturbations of the GAdSBH background.  A more efficient route is to replace \eno{zFoliator} by $\tilde{q}(y) = L/\rho$, but there is a minor complication, discussed in detail in \cite{Balasubramanian:1999re}: $\langle \tilde{T}_{ab} \rangle$ as obtained via \eno{BMetAndT}, using $\tilde{q}(y)$, now includes a term which accounts for the Casimir energy of the gauge theory on $S^3 \times {\bf R}$ with metric
 \eqn{NaturalSthreeMet}{
  d\tilde{s}^2 = \tilde{g}_{ab} dy^a dy^b =
    -d\tau^2 + L^2 d\hat\Omega^2 \,,
 }
where $d\hat\Omega^2$ is the metric on the unit $S^3$, as in \eno{sThreeMet}.  This Casimir term may be computed in a pure $AdS_5$ background and subtracted away: it is independent of the black hole.\footnote{Actually, it is sometimes convenient to make the subtraction work even harder: both the counterterms in \eno{BYtensor} may be dropped, and the corresponding divergences cancel in the course of the subtraction.}  Let the result of this subtraction be denoted $\langle \tilde{T}_{ab} \rangle_{\rm sub}$, and let
 \eqn{Wdef}{
  W = \lim_{\rho \to \infty} {q(y) \over \tilde{q}(y)} =
    {1 \over \cos {\tau \over L} + \cos\chi} =
   {1 \over 2L^2} \sqrt{4 t^2 L^2 + (L^2 + r^2 - t^2)^2} \,.
 }
Then, as a more efficient alternative to \eno{FinalT}, we may use
 \eqn{EfficientT}{
  \langle T_{mn} \rangle =
    {\partial y^a \over \partial x^m}
    {\partial y^b \over \partial x^n}
    {1 \over W^2} \langle \tilde{T}_{ab} \rangle_{\rm sub} \,.
 }
It is useful to note that
 \eqn{FindTauChi}{
  \sin {\tau \over L} = {t / L \over W} \qquad
   \sin\chi = {r/L \over W} \,,
 }
which follows from combining \eno{YtoX} and~\eno{Wdef}.

A natural expectation is that at large $\rho$, graviton perturbations  of GAdSBH take the form
 \eqn{LargeRhoForm}{
  \delta g_{AB} = {e^{-i\omega\tau} \over \rho^2}
    \eta_{AB}(\chi,\theta,\phi) + O(\rho^{-4}) \,.
 }
Because $\delta g_{AB}$ is a real field, \eno{LargeRhoForm} cannot be quite right for generic, complex $\omega$.  Instead, $\delta g_{AB}$ is a linear combination of terms of the form appearing on the right hand side, with the various coefficients arranged so that the sum is real.  The stress tensor is a sum of terms:
 \eqn{SplitT}{
  \langle \tilde{T}_{ab} \rangle_{\rm sub} =
    \langle \tilde{T}_{ab}^{\rm Sch} \rangle_{\rm sub} +
    \langle \tilde{T}_{ab}^{\rm QNM} \rangle_{\rm sub} \,,
 }
where the first term corresponds to the unperturbed GAdSBH background (see section~\ref{RADIAL} for an explicit evaluation) and the second accounts for the perturbations.  If we use \eno{LargeRhoForm} and ignoring the reality issues for the present, $\langle \tilde{T}_{ab}^{\rm QNM} \rangle_{\rm sub}$ is a linear expression in $\eta_{AB}$.

As a further simplification, consider the case where $\delta g_{\rho A} = 0$.  This is axial gauge, and it is always possible to pass to it using diffeomorphism gauge invariance.  Then using the foliator $\tilde{q}(y) = L/\rho$ in \eno{BYtensor} and \eno{BMetAndT}, one obtains
 \eqn{Tab}{
  \langle \tilde{T}_{ab}^{\rm QNM} \rangle_{\rm sub} =
    {2 \over \kappa^2 L^3} e^{-i\omega \tau}
     \left( \eta_{ab} - \tilde{g}^{cd} \eta_{cd} \tilde{g}_{ab}
    \right) \,,
 }
where $\tilde{g}_{ab}$ is the metric on $S^3 \times {\bf R}$ given in \eno{NaturalSthreeMet}.  The QNM contribution to the stress tensor $\langle T_{mn}^{\rm QNM} \rangle$ on ${\bf R}^{3, 1}$ can then be obtained from \eno{EfficientT}.

\subsection{The conformal soliton flow}
\label{RADIAL}

To understand quantitatively how the GAdSBH background translates into a radial flow in the boundary gauge theory, the main steps are to justify the choice \eno{zFoliator} of a foliating function for surfaces to which we then apply the holographic stress tensor construction.  To this end, let us start by describing global $AdS_5$ {\it without} a black hole: it is the covering space of the following hyperboloid in ${\bf R}^{4,2}$:
 \eqn{Xmanifold}{
  -(X^{-1})^2 - (X^0)^2 + (X^1)^2 + (X^2)^2 + (X^3)^2 +
    (X^4)^2 = -L^2 \,.
 }
The standard metric on $AdS_5$ is the one inherited from ${\bf R}^{4,2}$.  The embedding \eno{Xmanifold} is a convenient representation of $AdS_5$ because the action of the conformal group $SO(4,2)$ is obvious.\footnote{For a fuller account, see for example section 2.2.1 of \cite{MAGOO}.}  The relation to the coordinates used in \eno{GAdSBHmetric} (setting $\rho_0=0$) is
 \eqn{globalToY}{
  X^{-1} = L \sqrt{1 + {\rho^2 \over L^2}} \cos {\tau \over L} \qquad
   X^0 = L \sqrt{1 + {\rho^2 \over L^2}} \sin {\tau \over L} \qquad
   X^i = \rho \Omega_i
 }
where $\Omega_i$ are the components of a unit vector in ${\bf R}^4$.  Poincar\'e coordinates are the $(t,\vec{x},z)$ coordinates used to represent $AdS_5$ in the following form:
 \eqn{StandardAdS}{
  ds^2 = {L^2 \over z^2} (-dt^2 + d\vec{x}^2 + dz^2) \,.
 }
These coordinates cover only part of the hyperboloid and can be defined so that
 \eqn[c]{globalToX}{
  X^{-1} = {z \over 2} \left( 1 + {L^2+\vec{x}^2 - t^2 \over z^2}
    \right) \qquad  X^0 = L {t \over z}  \cr
  X^i = L {x^i \over z} \qquad
   X^4 = {z \over 2} \left( -1 + {L^2-\vec{x}^2 + t^2 \over z^2}
    \right) \,,
 }
where in the third equality, $i$ runs from $1$ to $3$.

The $X^M$ can be eliminated to obtain coordinate transformations from global to Poincar\'e coordinates.  To obtain explicit formulas it helps to express
 \eqn{OmegaIexpress}{
  \Omega_i = (\sin\chi \, \vec\Omega,\,\cos\chi) \qquad
   \vec\Omega = (\sin\theta \cos\phi,\, \sin\theta \sin\phi,\,
     \cos\theta) \,.
 }
$\vec\Omega$ is a unit vector in ${\bf R}^3$.  Then
 \eqn{globalToPoincare}{
  {z \over L} = {1 \over \sqrt{1+\rho^2/L^2} \cos {\tau \over L} +
    {\rho \over L} \cos\chi} \qquad
   t = z \sqrt{1+\rho^2/L^2} \sin {\tau \over L} \qquad
   \vec{x} = z {\rho \over L} \sin\chi \, \vec\Omega \,.
 }
To understand \eno{globalToPoincare}, it helps to consider its action on the boundary.  The boundary of global $AdS_5$ is ${\bf R} \times S^3$ (the Einstein static universe), while the boundary of the ``Poincar\'e patch'' covered by the coordinates $(t,\vec{x},z)$ is ${\bf R}^{3,1}$ (Minkowski space).  The $\rho \to \infty$ limit of \eno{globalToPoincare} defines an embedding of Minkowski space in the Einstein static universe:
 \eqn{EmbedMinkowski}{
  {t \over L} = {\sin{\tau \over L} \over
    \cos{\tau \over L} + \cos\chi} \qquad
   {\vec{x} \over L} = {\sin\chi \over \cos{\tau \over L} + \cos\chi}
     \vec\Omega \,.
 }
The Minkowski metric is related by a conformal transformation to the standard metric on the Einstein static universe:
 \eqn{MinkowskiTransforms}{
  ds^2 = -dt^2 + dr^2 + r^2 d\vec\Omega^2 =
    W^2 (-d\tau^2 + L^2 d\hat\Omega^2)
   \qquad W = {1 \over \cos{\tau \over L} + \cos\chi}
 }
where $d\vec\Omega^2$ is the metric on $S^2$ and $d\hat\Omega^2 = d\chi^2 + \sin^2 \chi d\vec\Omega^2$ is the metric on $S^3$, parametrized just as in \eno{sThreeMet}.  $W$ is the conformal factor.  Note that Minkowski space covers the patch of the Einstein static universe where $|\tau/L| + |\chi| < \pi$ (i.e.~where $1/W = \cos{\tau \over L} + \cos\chi$ is positive).

The coordinate transformation \eno{globalToPoincare} may be applied to the GAdSBH background \eno{GAdSBHmetric}.  The result is a black hole which is closest to the boundary (in the sense of attaining the smallest value of $z$) at time $t=0$.  For $t>0$, the black hole ``falls'' toward larger and larger values of $z$, and eventually it passes out of the Poincar\'e patch and is no longer causally accessible to the Minkowski space patch of the boundary on which we wish to calculate $\langle T_{mn} \rangle$.  The metric for this ``falling'' black hole, though entirely equivalent to \eno{GAdSBHmetric}, has a complicated form, and we will not reproduce it here.  In order to evaluate $\langle T_{mn} \rangle$, what one does in principle is to expand this complicated metric around the pure $AdS_5$ background \eno{StandardAdS} for small $z$ and extract the terms that fall to zero as $z^2$ for small $z$.  $\langle T_{mn} \rangle$ is expressed as a linear combination of the coefficients of these terms: see for example (40) of \cite{Friess:2006fk}.  In practice, because of the complicated form of the metric in Poincar\'e coordinates, we prefer to work with the global coordinates until the last step, using \eno{FinalT} or \eno{EfficientT}.  We now recognize \eno{zFoliator} as $z/L$, which means that the surfaces $\Sigma(\epsilon)$ on which the Brown-York quasi-local stress tensor is evaluated are, appropriately, surfaces of constant $z$ (sometimes called horospheres).

Before turning to the computation of $\langle T_{mn} \rangle$, let us detour to an explanation of the four depictions of the GAdSBH background in figure~\ref{GlobalBH}:
 \begin{figure}
  \centerline{\includegraphics[width=6.5in]{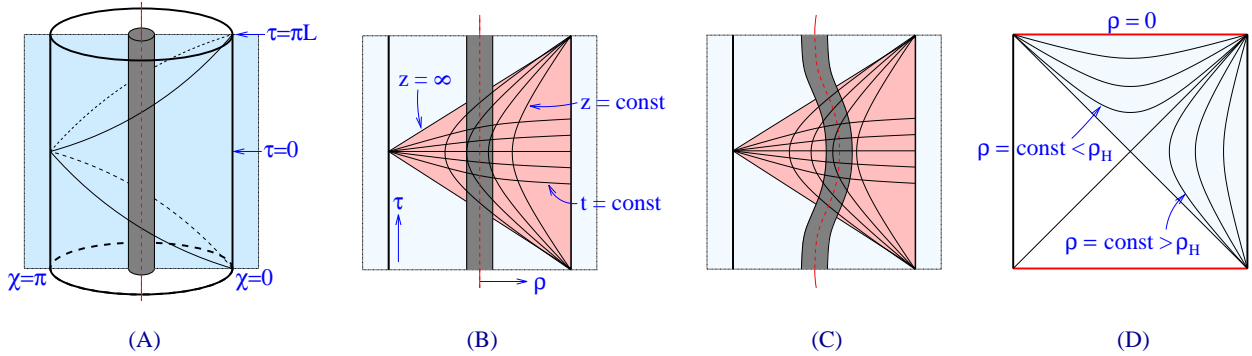}}
  \caption{
  Four complementary views of the GAdSBH background \eno{GAdSBHmetric}.  The first three are cartoons based on the Penrose diagram for global $AdS_5$.  The fourth is a true Penrose diagram of GAdSBH.  See the main text for more detailed explanations.}\label{GlobalBH}
 \end{figure}
 \begin{enumerate}
  \item[(A)] The boundary at $\rho=\infty$ is shown as a cylinder.  Actually it is the Einstein static universe, ${\bf R} \times S^3$.  The wedge shaped slice is the Poincar\'e patch.  The black hole passes into and then back out of the Poincar\'e patch.  The thin red line is the singularity at $\rho=0$.  The blue plane cuts the cylinder at $\chi = 0\,\hbox{mod}\;\pi$.
  \item[(B)] The cross section of figure~(A) is shown, with the Poincar\'e patch shaded in pink.  Surfaces of constant $z$ and constant $t$ are indicated.  Note that the black hole reaches its minimum $z$ at $t=0$.
  \item[(C)] A conformal transformation of the standard $AdS_5$-Schwarzschild metric has the black hole wandering through $AdS_5$.  The particular trajectory shown corresponds to a uniform dilation in the boundary theory.
  \item[(D)] The Penrose diagram of the GAdSBH geometry.  The region covered by the coordinates $(\tau,\rho)$ is shaded light blue, and corresponds to the light blue regions in the other figures.  The black hole singularity at $\rho=0$ is the upper red line.  The right boundary corresponds to $\rho=\infty$, and the left boundary is another copy of ${\bf R} \times S^3$ causally separated from the one at $\rho=\infty$.
 \end{enumerate}

Choosing the foliator $\tilde{q}(y) = L/\rho$ as described in the paragraph after \eno{FinalT}, and using \eno{BMetAndT} with $q$ replaced by $\tilde{q}$, one may straightforwardly show that
 \eqn{FoundSthreeT}{
  \langle \tilde{T}^{\rm Sch}_{ab} \rangle_{\rm sub} =
  {\rho_0^2 \over 2\kappa^2 L}
    \diag\left\{ {3 \over L^2},\;1,\;\sin^2\chi,\;
      \sin^2\chi \sin^2\theta \right\} \,.
 }
What makes this computation relatively straightforward is that the extrinsic curvature is simply
 \eqn{SimpleKAB}{
  K_{AB} = -{1 \over 2 \sqrt{g_{\rho\rho}}} \partial_\rho h_{AB} \,.
 }
The formula \eno{SimpleKAB} holds because of the vanishing of the shift function for the metric \eno{GAdSBHmetric} expressed in ADM form with respect to the foliation by surfaces of constant $\rho$.

As noted previously, there is a $\rho_0$-independent Casimir contribution to the stress tensor that has been subtracted away from the result quoted in \eno{FoundBackgroundT}.  \eno{FoundSthreeT} is the expected result that the GAdSBH metric is dual to a stationary plasma on $S^3$.  Using \eno{LkappaRelation} and \eno{EfficientT} we immediately extract the stress tensor on Minkowski space:
 \eqn{FoundBackgroundT}{
  \langle T^{\rm Sch}_{mn} \rangle = {\rho_0^2 \over 2 L^{10} W^4}
    \left( {N \over 2\pi} \right)^2
   \begin{pmatrix}
    3 L^4 + 4 {t^2 r^2 \over W^2} &
      -2 {t r \over W^2} (L^2 + t^2 + r^2) & 0 & 0  \\
    -2 {t r \over W^2} (L^2 + t^2 + r^2) &
      L^4 + 4 {t^2 r^2 \over W^2} & 0 & 0  \\
    0 & 0 & r^2 L^4 & 0  \\
    0 & 0 & 0 & r^2 L^4 \sin^2\theta
   \end{pmatrix} \,.
 }
Note that we are not considering in \eno{FoundBackgroundT} the generalization where an $SO(4,2)$ rotation corresponding to a dilation on the boundary is applied to the asymptotically $AdS_5$ spacetime, as in figure~\ref{GlobalBH}c.  Such a dilation can be applied after the final answer (e.g.~\eno{FoundBackgroundT} or a perturbation of it) is obtained, and it amounts, roughly speaking, to changing $L$ in \eno{FoundBackgroundT}.

As a check, we have also obtained \eno{FoundBackgroundT} more directly using the foliator $q(y) = z/L$ and the formula \eno{FinalT}.

Some further observations about this solution may be of interest:
 \begin{enumerate}
  \item There is no shear in the flow \eno{FoundBackgroundT}.
  \item The same solution exists in free conformal field theories, such as $U(1)$ gauge theory.
  \item All the properties of the flow are due to its conformal relation to a configuration on ${\bf R} \times S^3$ with constant energy density and pressure.  So its description as the ``conformal soliton flow'' seems appropriate.\label{ItsTrivial}
 \end{enumerate}

\subsection{Estimating the size of the black hole}
\label{SIZE}

Aside from new solutions generated from old ones through an $SO(4,2)$ transformation, there is precisely a one-parameter family of ``conformal solitons'' of the form \eno{FoundBackgroundT}: the parameter is $\rho_0$.  It is more interesting to cast this parameter in a form that is manifestly invariant under dilations.  Consider therefore the energy profile at time $t=0$:
 \eqn{epsilonZero}{
  \epsilon(0,r) = {24 L^2 \rho_0^2 \over (r^2+L^2)^4}
    \left( {N \over 2\pi} \right)^2 \,.
 }
It is easily seen that the total energy is
 \eqn{EtotDef}{
  E_{\rm tot} = {3 N^2 \rho_0^2 \over 4 L^3}
 }
and that the average of $r^2$ over the probability distribution $\epsilon(0,r) / E_{\rm tot}$ is $L^2$.  A dimensionless quantity---and therefore one invariant under dilations---is the combination
 \eqn{ELdef}{
  \Xi \equiv E_{\rm tot} L = {3N^2 \over 4} {\rho_0^2 \over L^2} \,.
 }
In our development through \eno{FoundBackgroundT}, $L$ is not an arbitrary length, but instead related to the five-dimensional gravitational constant through \eno{LkappaRelation}.  But in light of the dilation symmetry, we can regard $L$ as arbitrary in \eno{ELdef}.

It is easy to argue that $\Xi$ should be big in order to make a comparison to heavy ion collisions.  For example, without thinking too hard, we might take the following values for a central collision at RHIC:
 \eqn{RHICcomparison}{
  L = 7\,{\rm fm} \qquad E_{\rm tot} = 28\,{\rm TeV}
   \qquad\hbox{so}\quad
   \Xi \approx 10^6 \,.
 }
Using \eno{ELdef} with $N^2 = 8$ (the number of different gluons when the gauge group is $SU(3)$), one finds
 \eqn{EstimateRho}{
  {\rho_0 \over L} = \sqrt{4 \Xi \over 3 N^2} \approx 400 \qquad
   {\rho_H \over L} \approx 20 \,.
 }
There are some reasons to think that \eno{RHICcomparison} represents an over-estimate of the most appropriate values of $\Xi$.  The initial state in a Landau model of the collision is a Lorentz-flattened pancake whose long axes are the extent of the gold ions (so, a radius of about $7\,{\rm fm}$, which motivated the choice of $L$ in \eno{RHICcomparison}) and whose thickness is $100$ times smaller.  So should we decrease $\Xi$ by a factor of $100$?  Maybe not by so much, because the initial state is not thermalized.  If it thermalizes at $t \sim 1\,{\rm fm}/c$, then the aspect ratio of the region containing the QGP is more like $10$ than $100$, and perhaps $\Xi \sim 10^5$ is appropriate.  But we should also consider whether the value of $E_{\rm tot}$ in \eno{RHICcomparison} is appropriate.  This figure comes from an estimate that $28\,{\rm TeV}$ out of the initial $39\,{\rm TeV}$ in a gold-gold collision goes into heating and collective motion.  But it is not at all clear that all this energy thermalizes.  A conservative estimate of the thermalized energy would be to add up (as scalars) the transverse momenta of all the observed particles.  This is only about $4\,{\rm TeV}$.  In quoting these numbers we have relied on the summary discussions in \cite{Muller:2006ee}, which includes references to the primary literature.

The upshot of the considerations of the previous paragraph is that $\Xi$ is between $10^4$ and $10^6$, with the precise value depending on some assumptions.  The corresponding range of $\rho_H/L$, which is the parameter that will enter most frequently into QNM calculations, is $6$ to $20$.  Let us understand in another way what this parameter means.  The peak energy density in the conformal soliton flow is
 \eqn{ePeak}{
  \epsilon_{\rm peak} = {6N^2 \over \pi^2} {\rho_0^2 \over L^6} \,.
 }
Using the standard relations \cite{Gubser:1996de} for a plasma of infinite extent,
 \eqn{eosDthree}{
  {\epsilon \over 3} = p = {\pi^2 \over 8} N^2 T^4 \,,
 }
one extracts a peak temperature $T_{\rm peak}$ that satisfies
 \eqn{TPeak}{
  T_{\rm peak} L = {2 \over \pi} \sqrt{\rho_0 \over L}
     \approx {2 \over \pi} {\rho_H \over L} \,.
 }
Thus $\rho_H/L \gg 1$ is equivalent to $T_{\rm peak} L \gg 1$, which seems like a good assumption if the QGP thermalizes, since it is roughly the same as the condition that the Debye length should be much less than the extent of the plasma.  A good feature of \eno{TPeak} is that the dependence on $N$ cancels out.  In fact, a overall factor could be added to the relation \eno{LkappaRelation} between $\kappa^2/L^3$ and $N$, and \eno{TPeak} would not be affected.  This is because \eno{LkappaRelation} is used in the same way to get \eno{ePeak} and \eno{eosDthree}.  In other words, \eno{TPeak} doesn't depend on the heat capacity of the boundary gauge theory, which is good because the heat capacity for ${\cal N}=4$ super-Yang-Mills theory with gauge group $SU(3)$ is about three times bigger than for QCD.\footnote{The heat capacity is proportional to $g_*$ entering into the Stefan-Boltzmann relation $\epsilon_{\rm SB} = g_* {\pi^2 \over 30} T^4$.  In the weak coupling limit, $g_* = 120$ for $SU(3)$ ${\cal N}=4$, compared to $g_* = 37$ for QCD with two flavors and $g_* = 47.5$ for QCD with three flavors.  Lattice results show $g_* \approx 33$ for QCD with $2+1$ flavors at $1.5 T_c$, whereas in the limit of strong coupling, $g_* = 90$ in $SU(3)$ ${\cal N}=4$ if one uses the standard $3/4$ factor.}

\section{Quasinormal modes of global $AdS_5$-Schwarzschild}
\label{QUASINORMAL}

As we have remarked, the simplest gauge theory interpretation of the GAdSBH background is a static thermal plasma on $S^3 \times {\bf R}$.  Small fluctuations in the plasma damp out exponentially in global time $\tau$, and quasinormal modes provide a dual description of such behavior.  Readers familiar with \cite{Kovtun:2005ev} will note significant similarities to the results presented in this section.  This is no accident: the $S^3$ in which the nearly static plasma resides is much larger than the plasma's reciprocal temperature, so replacing it by ${\bf R}^3$ is at least qualitatively a good approximation.  However, for understanding finite-sized, expanding plasmas it is crucial to take the additional step of conformally transforming from $S^3 \times {\bf R}$ to Minkowski space.  This introduces additional time-dependence; or, more precisely, time-dependence is expressed in terms of the usual Minkowski time $t$ with respect to which the unperturbed background is not stationary.

\subsection{Tensor modes}
\label{TENSOR}

In section~\ref{TENSORFREQS} we explain how to numerically extract the frequencies of quasinormal modes from the master equation \eno{TensorEOM}.  In section~\ref{TENSORSTRESS} we show how tensor QNM's are translated into traceless, conserved perturbations of the conformal soliton flow \eno{FoundBackgroundT}.

There is significant overlap among the treatments of tensor, vector, and scalar modes.  In most ways the tensor modes are the simplest.  Common technical aspects are discussed more fully here than in sections~\ref{VECTOR} and~\ref{SCALAR}.

\subsubsection{Quasinormal frequencies}
\label{TENSORFREQS}

Using $f \approx \rho^2/L^2$ and $V_T \approx {15 \rho^2 / 4 L^4}$ for large $\rho$, and assuming $e^{-i\omega\tau}$ dependence, one finds the approximate form of the master equation \eno{TensorEOM}:
 \eqn{TeomFar}{
  \left( -\partial_\tau^2 +
    {\rho^2 \over L^4} \partial_\rho \rho^2 \partial_\rho -
    {15 \rho^2 \over 4 L^4} \right) \Phi_{\rm far} = 0 \,,
 }
which is solved by Bessel functions.  The allowed solution is
 \eqn{PhiFarTensor}{
  \Phi_{\rm far} = e^{-i\omega\tau} {8 \over L^4 \omega^2 \sqrt\rho}
    J_2(L^2 \omega / \rho) \approx e^{-i\omega\tau} \rho^{-5/2}
 }
for large $\rho$.

It is standard to introduce a ``tortoise'' coordinate
 \eqn{Tortoise}{
  r_* = \int {d\rho \over f(\rho)} =
    -{\sqrt{L^2+\rho_H^2} \tan^{-1} {\sqrt{L^2+\rho_H^2} \over
      \rho} + \rho_H \tanh^{-1} {\rho_H \over \rho} \over
    1 + 2\rho_H^2/L^2}
 }
which brings the master equation into the form
 \eqn{TensorSchrodinger}{
  \left[ -{\partial^2 \over \partial\tau^2} +
    {\partial^2 \over \partial r_*^2} - V_T(\rho) \right]
     \Phi = 0 \,.
 }
Because $V_T(\rho) \to 0$ as $\rho \to \rho_H$, one finds immediately that the allowed solution is
 \eqn{PhiNearTensor}{
  \Phi_{\rm near} = e^{-i\omega (\tau+r_*)} \,.
 }
$\Phi = e^{-i\omega (\tau-r_*)}$ is also a solution, but \eno{PhiNearTensor} is selected by requiring that $\Phi$ near the horizon is a function only of $\tau+r_*$, corresponding to a wave falling into the horizon.

In the spirit of factoring out the near-horizon behavior \eno{PhiNearTensor}, let us introduce a new radial variable, a rescaled frequency, and a separated ansatz for $\Phi$:
 \eqn{yPsiDef}{
  y = 1 - {\rho_H \over \rho} \qquad \Omega = \omega L \qquad
   \Phi = e^{-i\omega(\tau + r_*)} \psi(y) \,.
 }
The differential equation for $\psi(y)$ takes the form
 \eqn{psiEOM}{
  \left[ s(y) \partial_y^2 + t(y) \partial_y + u(y) \right]
    \psi(y) = 0
 }
where
 \eqn{stuDef}{
  s(y) &= K(y) (1-y)^4 f^2  \cr
  t(y) &= K(y) \left[ (1-y)^2 f {\partial \over \partial y}
    \left[ (1-y)^2 f \right] - 2i\omega \rho_H (1-y)^2 f \right]  \cr
  u(y) &= -K(y) \rho_H^2 V_T \,.
 }
Here $K(y)$ is chosen so that $s(y)$, $t(y)$, and $u(y)$ are polynomials in $y$ that do not all vanish at any given $y$.  The function $s(y)$ has a simple root at the horizon, $y=0$.  Because $y$ and $k_T$ are dimensionless, it must be possible to choose $K(y)$ so that $s$, $t$, and $u$ depend on $\rho_H$ and $L$ only through the dimensionless combination $\rho_H/L$.  We may expand
 \eqn{stuExpand}{
  s(y) = \sum_{i=1}^\infty s_i y^i \qquad
  t(y) = \sum_{i=0}^\infty t_i y^i \qquad
  u(y) = \sum_{i=0}^\infty u_i y^i \,,
 }
where of course all but finitely many of $s_n$, $t_n$, and $u_n$ vanish.  A formal series expansion of the solution $\psi(y)$ that approaches $1$ at the horizon may be developed as follows:
 \eqn[c]{psiExpand}{
  \psi(y) = \sum_{i=0}^\infty a_i y^i \qquad a_0 = 1  \cr
  a_j = {-1 \over j(j-1) s_1 + j t_0}
    \sum_{i=0}^{j-1} \left[ i(i-1) s_{j+1-i} +
      i t_{j-i} + u_{j-1-i} \right] a_i \qquad
   j > 0 \,.
 }

The strategy to determine $\Omega$ is to demand that a series solution for $\psi(y)$ around $y=0$ should converge at $y=1$.  This is valid provided there are no singularities of $\psi(y)$ closer to $y=0$ than $y=1$, which is true if $s(y)$ has no zeroes closer to $y=0$ than at $y=1$.  The zeroes of $s(y)$ are easily found: up to an arbitrary normalization,
 \eqn{sDef}{
  s(y) = y (1-y)^2 (2-y)^2
    \left( 1 + {i \rho_H^2 \over \rho_0 L} - y \right)^2
    \left( 1 - {i \rho_H^2 \over \rho_0 L} - y \right)^2 \,.
 }
Expressions for $t(y)$ and $u(y)$ are more complicated, and we do not gain much intuition from examining their explicit forms.  It is worth noting, however, that $s(y)$, $t(y)$, and $u(y)$ are all real when $\Omega$ is purely imaginary.  Because of this fact, $-\Omega^*$ is a quasinormal frequency if $\Omega$ is.  The boundary condition $\Phi \sim e^{i\Omega\tau/L} \rho^{-5/2}$ for large $\rho$ implies $\psi(y) \sim (1-y)^{5/2}$ for $y$ close to $1$, which is in contrast to the other possible boundary condition for metric perturbations, $\psi(y) \sim (1-y)^{-3/2}$.  So if we use $\psi(y,\Omega)$ to denote the solution of \eno{psiExpand} for a specified $\Omega$, the boundary condition becomes $\psi(1,\Omega) = 0$.  This transcendental equation determines the quasinormal frequencies $\Omega$.

To actually determine approximate numerical values for quasinormal frequencies, one may simply truncate the series solution \eno{psiExpand} at some number $M$ of terms.  The resulting function $\psi_M(y,\Omega)$ is a polynomial in $y$ and a rational expression in $\Omega$.  So solving $\psi_M(1,\Omega) = 0$ amounts to finding the roots of a polynomial in $\Omega$.  Unfortunately, for even moderate values of $M$ (such as $10$), the polynomials in question, when expressed analytically in terms of $\rho_H/L$, $k_T^2$, and $\Omega$, become large and cumbersome.  We have therefore found it more efficient to numerically evaluate $\psi_M(1,\Omega)$ on a grid of points in the complex $\Omega$ plane and use these values to scan for zeroes.  The results of the scans we performed in this way are summarized in table~\ref{TensorFreqTable}.  Note that the dependence on the $SO(4)$ quantum number $n$ is fairly weak, as is the dependence on $\rho_H/L$.  For each complex quasinormal frequency $\Omega$, there is of course another, $-\Omega^*$, as required by the reality of the master equation.
\newcommand{\TV}[3]{$\displaystyle{\seqalign{\span\TC}{#1 \cr\noalign{\vskip-1\jot} #2 \cr\noalign{\vskip-1\jot} #3}}$}
 \begin{table}
  \begin{center}
  \begin{tabular}{|c||c|c|c|c|}
   \hline
   $\hbox{freq} \backslash n$ & 3 & 4 & 5 & 6  \\ \hline\hline
$\Omega_1/\rho_H$ & \TV{3.220-2.725i}{3.141-2.742i}{3.129-2.745i}
 & \TV{3.264-2.712i}{3.151-2.739i}{3.133-2.743i}
 & \TV{3.319-2.696i}{3.163-2.735i}{3.138-2.742i}
 & \TV{3.385-2.677i}{3.178-2.731i}{3.145-2.740i}
\\ \hline $\Omega_2/\rho_H$ & \TV{5.281-4.748i}{5.193-4.760i}{5.180-4.762i}
 & \TV{5.312-4.739i}{5.200-4.758i}{5.183-4.761i}
 & \TV{5.352-4.728i}{5.209-4.756i}{5.186-4.760i}
 & \TV{5.400-4.714i}{5.220-4.753i}{5.191-4.759i}
\\ \hline $\Omega_3/\rho_H$ & \TV{7.318-6.757i}{7.216-6.767i}{7.200-6.768i}
 & \TV{7.344-6.750i}{7.221-6.765i}{7.202-6.768i}
 & \TV{7.376-6.741i}{7.229-6.763i}{7.205-6.767i}
 & \TV{7.415-6.730i}{7.237-6.761i}{7.209-6.766i}
\\ \hline $\Omega_4/\rho_H$ & \TV{9.350-8.761i}{9.230-8.770i}{9.211-8.771i}
 & \TV{9.372-8.755i}{9.235-8.769i}{9.213-8.771i}
 & \TV{9.399-8.748i}{9.241-8.767i}{9.216-8.770i}
 & \TV{9.433-8.738i}{9.248-8.765i}{9.219-8.769i}

 \\ \hline
  \end{tabular}
  \end{center}
  \caption{Frequencies of tensor quasinormal modes in units where $L=1$.  In each cell of the table, the top value is for $\rho_H=6$, the middle one is for $\rho_H=13$, and the bottom one is for $\rho_H=20$.}\label{TensorFreqTable}
 \end{table}

\subsubsection{Tensor perturbations of the conformal soliton}
\label{TENSORSTRESS}

The aim of the present section is to show that for {\it any} value of $\Omega$, a tensor quasinormal mode translates via AdS/CFT to a traceless, conserved perturbation of the stress tensor $\langle T_{mn}^{\rm Sch} \rangle$ of the conformal soliton flow~\eno{FoundBackgroundT}.  This means that there is {\it a priori} a whole function's worth of freedom in deforming the conformal soliton in the direction of a given tensor spherical harmonic.  The remarkable claim that can made using AdS/CFT is that a very restricted set of functions---namely, those that can be expressed as a sum of terms of the form $e^{-i\Omega\tau/L}$ where $\Omega$ is one of the values found in the previous section---describes the possible behaviors of conformal solitons in ${\cal N}=4$ super-Yang-Mills.

The strategy is to use \eno{Tab} to extract the contribution $\langle \tilde{T}_{ab}^{\rm QNM} \rangle_{\rm sub}$ of the tensor QNM to the stress tensor on $S^3 \times {\bf R}$ with the natural metric $\tilde{g}_{ab}$ described in \eno{NaturalSthreeMet}, and then apply the conformal transformation \eno{EfficientT} to recover $\langle T_{mn}^{\rm QNM} \rangle$ in Minkowski space.  The full stress tensor for the perturbed flow is
 \eqn{FullStress}{
  \langle T_{mn} \rangle = \langle T_{mn}^{\rm Sch} \rangle +
    \langle T_{mn}^{\rm QNM} \rangle \,.
 }
It is from $\langle T_{mn} \rangle$ that we would read off quantities like energy density that are supposed to be compared with the corresponding quantities in real-world quark-gluon plasmas.

Starting from \eno{TensorPerturb} and~\eno{TensorPhi} with
 \eqn{PhiFarAgain}{
  \Phi = e^{-i\Omega\tau/L} \rho^{-5/2} \,,
 }
one find that the metric perturbations can be expressed as in \eno{LargeRhoForm} with the large $\rho$ asymptotics
 \eqn{etaTensor}{
  \eta_{\alpha\beta} = \eta_{\alpha i} = 0 \qquad
    \eta_{ij} = 2 \mathbb{T}_{ij} \,.
 }
Explicit forms for $\mathbb{T}_{ij}$ are given in~\eno{TnlmEven} and~\eno{TnlmOdd}.  The perturbations \eno{etaTensor} are automatically in axial gauge, $\eta_{\alpha\rho} = 0$.  The trace of $\eta_{ab}$ vanishes:
 \eqn{etaTraceTensor}{
  \tilde{g}^{ab} \eta_{ab} = {1 \over L^2} \hat{g}^{ij} \eta_{ij}
     = {2 e^{-i\Omega\tau/L} \over L^2} \mathbb{T}^j{}_j = 0 \,.
 }
(Recall that $\hat{g}_{ij}$ is given by \eno{sThreeMet} while $\tilde{g}_{ab}$ is given by \eno{NaturalSthreeMet}.)  So we may express \eno{Tab} more simply as
 \eqn{TabAgain}{
  \langle \tilde{T}_{ab}^{\rm QNM} \rangle =
    {2 \over \kappa^2 L^3} e^{-i\Omega\tau/L} \eta_{ab} \,.
 }
To verify that $\langle \tilde{T}_{ab}^{\rm QNM} \rangle$ is conserved, first note that the connection on the factorized geometry $S^3 \times {\bf R}$ is $\tilde\nabla_\tau = \partial_\tau$ and $\tilde\nabla_i = \hat\nabla_i$, where as before $\hat\nabla_i$ is the standard connection on the unit $S^3$.  Thus
 \eqn{ConservedGeneral}{
  \tilde\nabla^a \langle \tilde{T}_{ab}^{\rm QNM} \rangle =
   -\partial_\tau \langle \tilde{T}_{\tau b}^{\rm QNM} \rangle +
    {1 \over L^2} \hat\nabla^i
     \langle \tilde{T}_{ib}^{\rm QNM} \rangle \,.
 }
The first term vanishes because $\langle \tilde{T}_{\tau b}^{\rm QNM} \rangle = 0$.  The second term is proportional to $\hat\nabla^i \mathbb{T}_{ij}$, which also vanishes according to the definition \eno{DefineHSHtensor}.

Putting \eno{EfficientT}, \eno{etaTensor}, and \eno{TabAgain} together, one winds up with the following final formula for a tensor perturbation to the conformal soliton flow:
 \eqn{FinalTensor}{
  \langle T_{mn}^{\rm QNM} \rangle =
    {\partial y^a \over \partial x^m}
    {\partial y^b \over \partial x^n} {2 \over L^6 W^2}
      \left( {N \over 2\pi} \right)^2
     e^{-i\Omega\tau/L} \eta_{ab} \,,
 }
where $\Omega$ is any of the dimensionless complex quasi-normal frequencies found in section~\ref{TENSORFREQS}, and $\eta_{ab}$ is given as in \eno{etaTensor}.\footnote{A minor subtlety arises for the case where $\Omega$ is purely imaginary, to be discussed in section~\ref{VECTORSTRESS}.}  Note that, after plugging in formulas for $\mathbb{T}_{ij}$ from~\eno{TnlmEven} or~\eno{TnlmOdd} and using \eno{Wdef} and \eno{FindTauChi}, the components of $\langle T_{mn}^{\rm QNM} \rangle$ arising from \eno{FinalTensor} are explicit algebraic expressions in terms of $t$, $r$, and trigonometric functions of $\theta$ and $\phi$.  Unfortunately these expressions are fairly complicated even for the simplest perturbations, and we will not reproduce them here.

The alert reader may notice that what we demonstrated in \eno{ConservedGeneral} is not the conservation of $\langle T_{mn}^{\rm QNM} \rangle$ with respect to the standard, flat connection on Minkowski space, but instead the conservation of $\langle \tilde{T}_{ab} \rangle_{\rm sub}$ with respect to the Christoffel connection $\tilde\nabla_a$ on $S^3 \times {\bf R}$, with metric as given by \eno{NaturalSthreeMet}.  But, together with tracelessness, conservation carries over from one conformal frame to another: one may think of the power of $W$ in \eno{FinalTensor} as just the right factor to ensure that the conformally transformed stress tensor is conserved with respect to the Christoffel connection of the conformally transformed metric.

\subsection{Vector modes}
\label{VECTOR}

Vector quasinormal modes are qualitatively similar to tensor modes except that, in addition to a tower of modes that is nearly isospectral with the tensor quasinormal frequencies shown in table~\ref{TensorFreqTable}, there is a single low-lying mode whose rescaled frequency $\Omega=\omega L$ scales as $L/\rho_H$, rather than the usual $\rho_H/L$.  Discussion of technical aspects in common with the case of tensor QNM's will be abbreviated.

\subsubsection{Quasinormal frequencies}
\label{VECTORFREQS}

Using $f \approx \rho^2/L^2$ and $V_V \approx {3 \rho^2 / 4 L^4}$ for large $\rho$, and assuming $e^{-i\Omega\tau/L}$ dependence, one finds two asymptotic solutions to the master equation \eno{VectorEOM}, with $\rho^{1/2}$ and $\rho^{-3/2}$ behavior, respectively.  The former corresponds to metric deformations, which we disallow.  The latter may be developed as a power series:\footnote{As with tensor modes, \eno{VectorEOM} can be solved in terms of hypergeometric functions in the case of pure $AdS_5$, but we will not need this solution.}
 \eqn{PhiFarVector}{
  \Phi_{\rm far} = e^{-i\Omega\tau/L} \rho^{-3/2}
   \left[ 1 - {2-k_V^2+\Omega^2 \over 8} {L^2 \over \rho^2} +
    O(\rho^{-4}) \right] \,.
 }
Expansion of $\Phi$ near the horizon in a power series times an overall exponential  proceeds in a fashion identical to the treatment \eno{TeomFar}-\eno{psiExpand} for tensor modes, except that $V_T$ is replaced by $V_V$, which also vanishes at the horizon.  Even the functions $K(y)$, $s(y)$ and $t(y)$ are the same, though $u(y)$ and hence $\psi(y)$ are different.  Thus the series expansion method described below~\eno{psiExpand} is justified, and we used it to scan for quasinormal frequencies $\Omega = \omega L$ in the complex plane.

A notable feature of the vector quasinormal frequencies is the occurrence of a single low-lying mode for each allowed value of $k_V^2$, namely $k_V^2 = n^2-2$ for $n=3,4,5,\ldots$, with frequencies which appear to be a good fit to the following formula, similar to a formula found in \cite{Cardoso:2003cj} in the $AdS_4$ case:
 \eqn{LowQNMs}{
  \Omega \approx -{i \over 4} {L \over \rho_H} (n^2-4) \,.
 }
(See section~\ref{FITS} for a more systematic discussion of the best fit formulas to low-lying quasinormal modes.)  Actually, each $\Omega$ in \eno{LowQNMs} corresponds to a pair of $(n^2-1)$-dimensional irreducible representations of $O(4)$, filled out by choosing $\mathbb{V}_i$ to be one of the explicit forms given in \eno{VnlmEven} or~\eno{VnlmOdd}.

Continuing the formula \eno{LowQNMs} to $n=2$ would suggest that there are vector zero modes, forming an adjoint representation of $O(4)$.  This is obviously true: there are rotational moduli of the black hole, and they indeed fall in the adjoint of $O(4)$.  But, as the authors of \cite{Kodama:2003jz} note, the derivation of the master equation \eno{VectorPerturb} relies upon assuming $n \geq 3$.  For the special case $n=2$, a simpler treatment is possible and was given in \cite{Kodama:2000fa,Kodama:2003jz}.  The relevant Einstein equations may be cast as conservation of a ``field strength:''
 \eqn{FieldFab}{
  \partial_\gamma \left( \rho^4
    \epsilon^{\alpha\beta} F_{\alpha\beta} \right) = 0
   \qquad\hbox{where}\quad
  F_{\alpha\beta} = \rho \, \partial_\alpha {f_\beta \over \rho} -
    \rho \, \partial_\beta {f_\alpha \over \rho} \,.
 }
$F_{\alpha\beta}$ has nothing to do with the symmetric tensor $F_{\alpha\beta}$ entering into \eno{ScalarPhi}.  There is only one solution of \eno{FieldFab} consistent with axial gauge, $f_\rho = 0$, and the vanishing of $f_\tau$ for $\rho \to 0$ that is imposed on us by demanding that the metric on the boundary is not perturbed.  It is
 \eqn{fTauZeroMode}{
  f_\tau = {Q \over 4\rho^3} \,,
 }
where $Q$ is the angular momentum resulting from turning on the perturbation.  Unlike the $n>2$ cases, there is no tower of non-zero quasi-normal frequencies.  The zero-mode \eno{fTauZeroMode} is the only allowed perturbation in the $n=2$ sector.  Setting $Q=1$, one immediately obtains
 \eqn{etaZeroMode}{
  \eta_{\tau\tau} = 0 \qquad
  \eta_{\tau i} = {1 \over 4} \mathbb{V}_i \qquad
  \eta_{ij} = 0 \,.
 }

Besides the low-lying mode described in \eno{LowQNMs}, there is a main series of quasinormal frequencies for all modes with $n>2$.  Remarkably, these are nearly isospectral with the tensor modes in the cases that we explored numerically, namely $\rho_H=6$, $13$, and $20$ with $3 \leq n \leq 6$: see table \ref{VectorFreqTable}.
 \begin{table}
  \begin{center}
  \begin{tabular}{|c||c|c|c|c|c|}
   \hline
   $\hbox{freq} \backslash n$ & 2 & 3 & 4 & 5 & 6  \\ \hline\hline
$\rho_H \Omega_0$ & 0 & \TV{-1.255i}{-1.246i}{-1.237i}
 & \TV{-3.038i}{-3.002i}{-2.990i}
 & \TV{-5.369i}{-5.269i}{-5.247i}
 & \TV{-8.283i}{-8.053i}{-8.011i}
\\ \hline $\Omega_1/\rho_H$ & N/A & \TV{3.184-2.736i}{3.133-2.744i}{3.125-2.746i}
 & \TV{3.207-2.729i}{3.138-2.743i}{3.128-2.745i}
 & \TV{3.235-2.720i}{3.145-2.741i}{3.130-2.744i}
 & \TV{3.268-2.708i}{3.152-2.738i}{3.133-2.743i}
\\ \hline $\Omega_2/\rho_H$ & N/A & \TV{5.256-4.755i}{5.188-4.761i}{5.178-4.762i}
 & \TV{5.272-4.750i}{5.192-4.760i}{5.179-4.762i}
 & \TV{5.292-4.744i}{5.196-4.759i}{5.181-4.761i}
 & \TV{5.316-4.737i}{5.202-4.758i}{5.183-4.761i}
\\ \hline $\Omega_3/\rho_H$ & N/A & \TV{7.299-6.761i}{7.212-6.767i}{7.199-6.768i}
 & \TV{7.312-6.758i}{7.215-6.766i}{7.200-6.767i}
 & \TV{7.328-6.753i}{7.219-6.765i}{7.201-6.767i}
 & \TV{7.348-6.747i}{7.223-6.764i}{7.203-6.767i}
\\ \hline $\Omega_4/\rho_H$ & N/A & \TV{9.334-8.764i}{9.227-8.769i}{9.211-8.770i}
 & \TV{9.345-8.761i}{9.230-8.769i}{9.212-8.770i}
 & \TV{9.359-8.757i}{9.233-8.768i}{9.213-8.769i}
 & \TV{9.376-8.752i}{9.237-8.767i}{9.215-8.769i}

 \\ \hline
  \end{tabular}
  \end{center}
  \caption{Frequencies of vector quasinormal modes in units where $L=1$.  In each cell of the table, the top value is for $\rho_H=6$, the middle one is for $\rho_H=13$, and the bottom one is for $\rho_H=20$.  The low-lying frequencies, approximated by \eno{LowQNMs}, are shown as $\Omega_0$.  The $n=2$ column contains only the zero mode described in \eno{FieldFab}-\eno{etaZeroMode}.}\label{VectorFreqTable}
 \end{table}

\subsubsection{Vector perturbations of the conformal soliton}
\label{VECTORSTRESS}

In this section, we wish to show that the expression \eno{TabAgain} still applies, with suitably altered $\eta_{ab}$, and that the resulting stress-energy tensor is conserved.

Plugging \eno{PhiFarVector} into \eqref{VectorPhi} we get
 \eqn{FtrVector}{
  F_\tau &= {2-k_V^2+\Omega^2 \over 4} e^{-i\Omega\tau/L}
    {1 \over \rho^3} +
   O\left(\rho^{-4}\right) \qquad
  F_\rho = -i \Omega e^{-i\Omega \tau/L} {L \over \rho^4} +
   O\left(\rho^{-5}\right)  \,.
 }
The quantities $f_\alpha$ and $H_T$ are not separately gauge invariant.  Roughly speaking, gauge choice at the level we are interested in amounts to choosing $H_T$.  Axial gauge, $f_\rho = 0$, can be arranged by choosing
 \eqn{HTasymp}{
  H_T = i {\Omega k_V \over 4} e^{-i \Omega \tau/L}
    {L \over \rho^4} + O\left(\rho^{-5}\right) \,.
 }
Then
 \eqn{ftauVector}{
  f_\tau = {2-k_V^2 \over 4} e^{-i\Omega\tau/L}
    {1 \over \rho^3} + O(\rho^{-4})
 }
and
 \eqn{etaVector}{
  \eta_{\tau\tau} = 0 \qquad
  \eta_{\tau i} = {2-k_V^2 \over 4} \mathbb{V}_i \qquad
  \eta_{ij} = -i {L\Omega \over 2} \hat\nabla_{(i}
    \mathbb{V}_{j)} \,.
 }
The trace $\tilde{g}^{ab} \eta_{ab}$ is proportional to $\hat{g}^{ij} \hat\nabla_{(i} \mathbb{V}_{j)}$, which vanishes by the definition \eno{DefineHSHvector} of $\mathbb{V}_j$.  So we can again use the simplified relation \eno{TabAgain} for the stress tensor, and what remains is to check that $e^{-i\Omega\tau/L} \eta_{ab}$ is conserved.  For $b=\tau$ this is easy:
 \eqn{tVectorConserve}{
  \tilde\nabla^a \left( e^{-i\Omega\tau/L} \eta_{a\tau} \right)
    = {1 \over L^2} e^{-i\Omega\tau/L} \hat\nabla^i \eta_{i\tau} \,,
 }
and the right hand side vanishes because it is proportional to $\hat\nabla^i \mathbb{V}_i$.  For $b \neq \tau$ we have
 \eqn{TwoTerms}{
  \tilde\nabla^a \left(e^{-i\Omega\tau/L} \eta_{ai}\right) &=
    -\partial_\tau \left(e^{-i\Omega\tau/L} \eta_{\tau i}\right) +
    {1\over L^2} e^{-i\Omega\tau/L} \hat\nabla^j \eta_{ji}  \cr
   &= i {2-k_V^2 \over 4} {\Omega \over L} e^{-i\Omega\tau/L}
     \mathbb{V}_i -
    i {\Omega \over 4L} e^{-i\Omega\tau/L}
      \, 2 \hat\nabla^j \hat\nabla_{(i} \mathbb{V}_{j)} \,.
 }
To show that the last expression vanishes, we compute
 \eqn{ddV}{
  2 \hat\nabla^j \hat\nabla_{(i} \mathbb{V}_{j)} &\equiv
   \hat\nabla^j \hat\nabla_j \mathbb{V}_i +
    \hat\nabla^j \hat\nabla_i \mathbb{V}_j =
   -k_V^2 \mathbb{V}_i +
    [\hat\nabla^j,\hat\nabla_i] \mathbb{V}_j  \cr
   &= -k_V^2 \mathbb{V}_i - \hat{R}^j{}_i{}^k{}_j \mathbb{V}_k
   = -k_V^2 \mathbb{V}_i + \hat{R}_i{}^k \mathbb{V}_k
   = (2-k_V^2) \mathbb{V}_i \,,
 }
where in the last step we used $\hat{R}_{ij} = 2\hat{g}_{ij}$ for the unit $S^3$.

\subsection{Scalar modes}
\label{SCALAR}

Scalar modes are for several reasons more technically challenging than the tensor and vector cases:
 \begin{enumerate}
  \item The potential $V_S(\rho)$ from \eno{ScalarPotential} is more complicated, resulting in polynomials $s(y)$, $t(y)$, and $u(y)$ that are higher order than before.
  \item The far field behavior of $\psi(y)$ (defined analogously to \eno{yPsiDef}) is either $\sqrt{1-y}$ or $\sqrt{1-y} \times \log(1-y)$.
The latter is disallowed by the usual conditions of not changing the metric on the boundary, but it is numerically challenging to tell the difference between functions with these two behaviors.  In particular, we can no longer simply evaluate $\psi_N(y)$ at $y=1$ and look for the values of $\Omega$ where it vanishes: with sufficient accuracy, it will vanish for all $\Omega$!
  \item There are two nearly degenerate quasinormal frequencies close to the origin, and it is a challenge to obtain enough numerical resolving power to distinguish them.
 \end{enumerate}
As a final illustration of the challenges, we note that the methods of sections~\ref{VECTOR} and~\ref{TENSOR}, applied without regard to point~2 and carried to $120$ terms in the series and $150$ digits of working precision, sometimes gave values of $\Omega$ with $\Im\Omega$ significantly greater than $0$, contracting the stability proof of \cite{Ishibashi:2003ap}.

\begin{figure}
  \centerline{\includegraphics[width=4in]{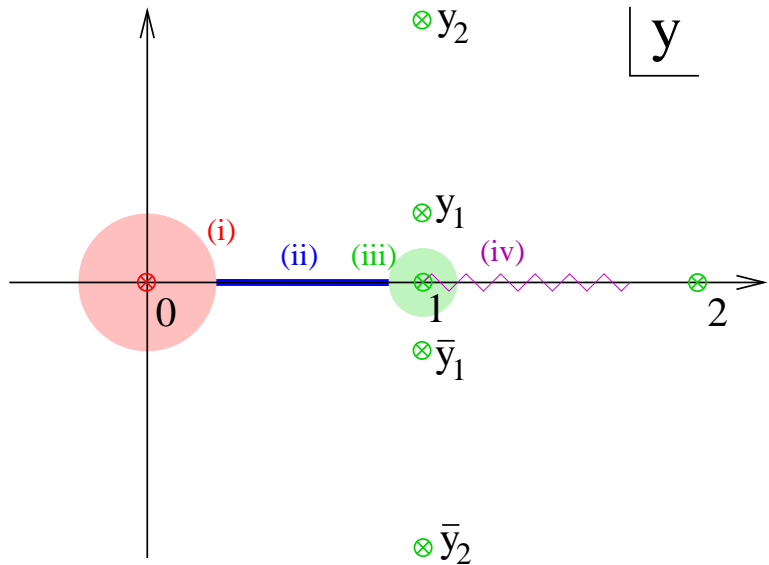}}
  \caption{The zeroes of $s(y)$ are shown as crosses in the complex $y$ plane.  The red cross at the origin is a single zero; all others are double zeroes.  The region (i) is where a power series expansion around $y=0$ was used in computing $\psi(y)$.  The line (ii) is where $\psi(y)$ was computed by numerical integration, seeded with initial conditions at the boundary with region (i).  The region (iii) is where $\psi(y)$ was computed using a power series around $y=1$.  The powers in this latter series are half-integral, so there is a branch cut which may be located as shown by (iv).}\label{YPlaneS}
 \end{figure}

To surmount the difficulties just described, we adopted a modified approach illustrated in figure~\ref{YPlaneS}.
The plan is to use numerical integration of the differential equation \eno{psiEOM} (implemented through Mathematica's {\tt NDSolve}) to compute $\psi(y)$ from initial conditions at $y=y_i=1/4$ obtained through a power series expansion of the form \eno{psiExpand} around $y=0$ up to a value $y=y_f$ that is half as far from $y=1$ as the zeroes $y=y_1$ and $y=\bar{y}_1$.  In place of $\psi(1)$ we computed the Wronskian between the numerically evaluated $\psi(y)$ and the series expansion of the allowed solution around $y=1$.  This Wronskian vanishes only if the same $\psi(y)$ that has regular behavior at the horizon has $\sqrt{1-y}$ behavior near $y=1$: in other words, for quasinormal modes obeying standard AdS/CFT boundary conditions.

The rationale for this plan is that $\psi(y)$ must be analytic away from the zeroes of $s(y)$ and (sometimes) branch cuts emanating from these zeroes.  Up to an arbitrary normalization,
 \eqn[c]{sExpress}{
  s(y) = y (y-1)^2 (y-2)^2 (y-y_1)^2 (y-\bar{y}_1)^2
    (y-y_2)^2 (y-\bar{y}_2)^2  \cr
  y_1 = 1 + {i \over \sqrt{6}} \sqrt{k_S^2-3 \over 1+\rho_H^2}
     \qquad
  y_2 = 1 + {i \rho_H \over \sqrt{1+\rho_H^2}} \,.
 }
Further details of the computation of quasinormal modes are presented in section~\ref{SCALARFREQS}, followed by a discussion in section~\ref{SCALARSTRESS} of the dual stress tensor perturbations of the conformal soliton flow.

\subsubsection{Quasinormal frequencies}
\label{SCALARFREQS}

From the large $\rho$ asymptotics $f \approx \rho^2/L^2$, $V_S \approx - \rho^2/4 L^4$, and the assumption $\Phi \propto e^{-i \Omega \tau/L}$, one finds that the two linearly independent solutions of \eno{ScalarEOM} behave asymptotically as $\rho^{-1/2} \log \rho$ and $\rho^{-1/2}$ at large $\rho$, corresponding to the $\sqrt{1-y} \log(1-y)$ and $\sqrt{1-y}$ behaviors for $\psi(y)$ noted previously.  The former solution should be disregarded because it corresponds to deformations of the boundary metric.  In section~\ref{SCALARFREQS} we will have occasion to use the series expansion
 \eqn{PhiFarScalar}{
  \Phi_{\rm far} = e^{-i \Omega \tau/L} \rho^{-1/2} \Bigg[1 + \left(-{6 \rho_0^2 \over m L^2} + {k_S^2 - \Omega^2 \over 4} \right) {L^2 \over \rho^2} + \bigg({m^2 \over 64} + {36 \rho_0^4 \over m^2 L^4} - {m (1 + \Omega^2) \over 32} \cr
   {}+{3 \rho_0^2 (-5 + \Omega^2) \over 4 m L^2} - {80 \rho_0^2 +L^2 (15 - 6 \Omega^2 - \Omega^4) \over 64 L^2} \bigg) {L^4\over \rho^4} + O\left(\rho^{-6}\right) \Bigg]\,,
 }
where $m = k_S^2-3$, as in \eno{ScalarParams}.

After defining $\psi(y)$ as in \eno{yPsiDef}, a treatment entirely analogous to \eno{psiExpand} is possible near the horizon.  A significant difficulty is that the recursion relations involve many terms, so that evaluating $a_j$ symbolically is impractical for $j$ greater than a few.  Numerical evaluations of the $a_j$ lose accuracy as $j$ increases on account of the round-off error arising from iterated evaluations of the recursion relations.  Experience shows that high accuracy can easily be obtained from the power series expansion up to $y$ of $1/4$, or even $1/2$.

The series expansion of the allowed $\sqrt{1-y}$ solution around $y=1$ proceeds in a fashion rather similar to \eno{psiExpand}, and we omit the details.  Experience shows that high accuracy can easily be obtained from this series expansion out to $1/2$ or $2/3$ of the radius of convergence.  Unfortunately, as illustrated in figure~\ref{YPlaneS}, the two regions of good convergence do not overlap.  To cross the gap between these two regions, we resorted to numerical integration of the differential equation for $\psi$, of the form \eno{psiEOM} (but, obviously, with appropriately altered polynomials $s(y)$, $t(y)$, and $u(y)$).  As previously remarked, the function of $\Omega$ whose zeroes determine the quasinormal frequencies is, in our approach, the Wronskian at $y_f = 1-|1-y_1|/2$ between a numerically integrated form of $\psi(y)$ which obeys the usual horizon boundary conditions and the allowed series solution around $y=1$.

Similarly to the vector mode case, the scalar modes consist of low-lying modes and a main series.\footnote{We believe that the same pattern should persist for scalar modes (more commonly described as even modes) in $AdS_4$.  Main series modes were found in \cite{Cardoso:2003cj}, but the low-lying modes were not.  In section~\ref{HYDRO} we will comment on the hydrodynamical interpretation of the low-lying modes.  The hydrodynamical discussion should apply with small adaptations to the $AdS_4$ case.}  Remarkably, the main series turns out to be again nearly isospectral with the tensor and vector modes for the cases that we examined, namely for $\rho_H = 6$, $13$, and $20$, and $3 \leq n \leq 6$: see table \ref{ScalarFreqTable}.  The low-lying scalar modes differ from the vector ones in that they have non-zero real parts.  Their imaginary parts are roughly $2/3$ times the imaginary parts of the corresponding vector modes.
 \begin{table}
  \begin{center}
  \begin{tabular}{|c||c|c|c|c|}
   \hline
   $\hbox{freq} \backslash n$ & 3 & 4 & 5 & 6  \\ \hline\hline
$\re \Omega_0+ \rho_H \im \Omega_0$ & \
\TV{1.652-0.826i}{1.637-0.832i}{1.640-0.842i}
 & \TV{2.292-1.972i}{2.248-1.994i}{2.241-1.997i}
 & \TV{2.949-3.428i}{2.854-3.485i}{2.839-3.494i}
 & \TV{3.633-5.179i}{3.462-5.303i}{3.435-5.321i}
\\ \hline $\Omega_1/\rho_H$ & \TV{3.172-2.740i}{3.131-2.745i}{3.124-2.746i}
 & \TV{3.188-2.735i}{3.134-2.744i}{3.126-2.746i}
 & \TV{3.208-2.729i}{3.138-2.743i}{3.127-2.745i}
 & \TV{3.233-2.723i}{3.143-2.741i}{3.130-2.744i}
\\ \hline $\Omega_2/\rho_H$ & \TV{5.247-4.758i}{5.186-4.762i}{5.177-4.763i}
 & \TV{5.258-4.755i}{5.188-4.762i}{5.178-4.763i}
 & \TV{5.272-4.751i}{5.191-4.761i}{5.179-4.762i}
 & \TV{5.290-4.747i}{5.195-4.760i}{5.180-4.762i}
\\ \hline $\Omega_3/\rho_H$ & \TV{7.291-6.765i}{7.210-6.768i}{7.197-6.769i}
 & \TV{7.300-6.762i}{7.212-6.768i}{7.198-6.769i}
 & \TV{7.311-6.759i}{7.214-6.767i}{7.199-6.769i}
 & \TV{7.325-6.756i}{7.217-6.767i}{7.200-6.768i}
\\ \hline $\Omega_4/\rho_H$ & \TV{9.327-8.768i}{9.225-8.772i}{9.209-8.772i}
 & \TV{9.335-8.766i}{9.226-8.771i}{9.209-8.772i}
 & \TV{9.344-8.763i}{9.228-8.771i}{9.210-8.772i}
 & \TV{9.356-8.760i}{9.231-8.770i}{9.211-8.772i}

 \\ \hline
  \end{tabular}
  \end{center}
  \caption{Frequencies of scalar quasinormal modes in units where $L=1$.  In each cell of the table, the top value is for $\rho_H=6$, the middle one is for $\rho_H=13$, and the bottom one is for $\rho_H=20$.}\label{ScalarFreqTable}
 \end{table}

\subsubsection{Scalar perturbations of the conformal soliton}
\label{SCALARSTRESS}

Analogously to sections~\ref{TENSORSTRESS} and~\ref{VECTORSTRESS}, we wish to show in this section that \eno{TabAgain} can be used to compute the stress tensor dual to a scalar quasinormal mode, with suitably altered $\eta_{ab}$, and that this stress tensor is conserved.

Plugging \eno{PhiFarScalar} into \eno{ScalarPhi} gives
 \eqn{calF}{
  \mathcal{F}_{\tau \tau} &=  {k_S^2 - 3 \Omega^2 \over 4} e^{-i \Omega \tau/L} {1\over L^2}+ \mathcal{F}{1 \over \rho^2} + O\left(\rho^{-4}\right)\cr
  \mathcal{F}_{\rho \tau} &= -i \Omega {2 - k_S^2 + \Omega^2\over 2} e^{-i \Omega \tau/L} {L\over \rho^3} + O\left(\rho^{-5} \right) \cr
  \mathcal{F}_{\rho \rho} &= {k_S^2 - 3 \Omega^2 \over 4} e^{-i \Omega \tau/L} {L^2\over \rho^4} + \left(\mathcal{F} - {k_S^2 - 3 \Omega^2 \over 2} e^{-i \Omega \tau/L} \right) {L^4\over \rho^6} + O\left(\rho^{-8} \right)
 }
where
 \eqn{calFDef}{
  \mathcal{F} = {k_S^2 (3 k_S^2 - 8) - 10 \Omega^2 (k_S^2 - 2) + 7 \Omega^4 \over 32} e^{-i \Omega \tau/L} \,.
 }
We also note that \eno{ScalarPhi} implies the following relations between $F_{\alpha \beta}$ and $F$ (see \cite{Kodama:2003jz} for a more detailed explanation):
 \eqn{FRelations}{
  F_\alpha^{\phantom{\alpha}\alpha} + 2 F = 0 \qquad D^\beta \left(\rho F_{\alpha \beta} - 2 \rho F g_{\alpha \beta} \right) = 0\,.
 }

The above equations \eno{calF} and \eno{FRelations}, together with the axial gauge conditions $f_\rho = f_{\rho\rho} = f_{\tau \rho} = 0$, determine unambiguously the leading order dependence on $\rho$ in the metric perturbations \eqref{ScalarPerturb}.  Namely, $f_{\tau}$, $f_{\tau \tau}$, $H_L$, and $H_T$ are:
 \eqn{fandH}{
  f_\tau &= -{i \Omega k_S (k_S^2-3) \over 12} e^{-i \Omega \tau/L} {L \over \rho^3} + O\left(\rho^{-5}\right)\cr
  f_{\tau \tau} &= {k_S^2 (k_S^2-3)\over 12} e^{-i \Omega \tau/L} {1 \over \rho^2} + O\left(\rho^{-4}\right)\cr
  H_L &= {k_S^2 (k_S^2-3) \over 72} e^{-i \Omega \tau/L} {L^2 \over \rho^4} + O\left(\rho^{-6}\right)\cr
  H_T &= {k_S^2 (k_S^2-3 \Omega^2)\over 48} e^{-i \Omega \tau/L} {L^2 \over \rho^4} + O\left(\rho^{-6}\right)\,.
 }
When combined with \eqref{LargeRhoForm}, the above expressions give explicit formulas for $\eta_{ab}$:
 \eqn{etaScalar}{
  \eta_{\tau \tau} = {k_S^2 (k_S^2-3) \over 12} \mathbb{S} \qquad \eta_{\tau i} = -{i L \Omega k_S (k_S^2-3) \over 12}  \mathbb{S}_i\cr
  \eta_{ij} = {L^2 k_S^2 \over 36} \left[(k_S^2-3) \hat g_{ij} \mathbb{S} + {3 \over 2} (k_S^2-3 \Omega^2) \mathbb{S}_{ij} \right]\,,
 }
which in turn can be used to find the VEV of the QNM contribution to the stress tensor in the boundary theory via \eqref{Tab}.

Equation \eqref{Tab} contains the trace $\tilde g^{ab} \eta_{ab}$, which was zero for the tensor and vector modes.  Likewise, for the scalar modes, we obtain from \eno{etaScalar}
 \eqn{TraceScalar}{
  \tilde g^{ab} \eta_{ab} = - {k_S^2 (k_S^2-3) \over 12} \mathbb{S} +
    {k_S^2 \over 36} \left[ 3 (k_S^2-3) \mathbb{S} +
      {3 \over 2} (k_S^2-3\Omega^2) \mathbb{S}^i{}_i \right] = 0 \,.
 }
We have used $\mathbb{S}^i_{\phantom{i}i} = 0$, which follows from the definitions \eqref{ScalarSDefs} of $\mathbb{S}_{ij}$ and \eqref{DefineHSHscalar} of $\mathbb{S}$.  Therefore, $\eta$ is traceless in this case too, and we can again use the simplified equation \eqref{TabAgain} to compute the VEV of the stress tensor.

It remains to check that the holographic stress tensor is conserved, which is the same as checking that $\tilde \nabla^a \left(e^{-i \Omega \tau/L} \eta_{ab} \right) = 0$.  To do this, we will use the identities \cite{Kodama:2000fa}
 \eqn{SRelations}{
  \partial_i \mathbb{S} = - k_S \mathbb{S}_i \qquad
  \hat \nabla^i \mathbb{S}_i = k_S \mathbb{S} \qquad \hat \nabla^j \mathbb{S}_{ji} = {2 \over 3} {k_S^2 - 3 \over k_S} \mathbb{S}_i\,.
 }
The first of these is trivial, and the second and third follow in a more or less straightforward way from the definitions \eqref{DefineHSHscalar} and \eqref{ScalarSDefs} of $\mathbb{S}$, $\mathbb{S}_i$, and $\mathbb{S}_{ij}$.  To check that $\tilde\nabla^a (e^{-i \Omega \tau/L} \eta_{a\tau}) = 0$, we compute as follows:
 \eqn{tScalarConserv}{
  \tilde \nabla^a \left( e^{-i \Omega \tau/L} \eta_{a\tau} \right) &= -\partial_\tau \left( e^{-i \Omega \tau/L} \eta_{\tau \tau} \right) + {1\over L^2} \hat \nabla^i \left( e^{-i \Omega \tau/L} \eta_{i\tau} \right) \cr
  &= e^{-i \Omega \tau/L} \left[ i {\Omega \over L} {k_S^2 (k_S^2-3) \over 12} \mathbb{S} - {1\over L^2} {i L \Omega k_S (k_S^2-3)\over 12} \hat \nabla^i \mathbb{S}_i \right]\,,
 }
which vanishes by the second relation in \eqref{SRelations}.  Similarly,
 \eqn{iScalarConserv}{
  \tilde \nabla^a \left( e^{-i \Omega \tau/L} \eta_{ai} \right) &= -\partial_\tau \left( e^{-i \Omega \tau/L} \eta_{\tau i} \right) + {1\over L^2} \hat \nabla^j \left( e^{-i \Omega \tau/L} \eta_{ji} \right) \cr
  &= e^{-i \Omega \tau/L} \left[{\Omega^2 k_S (k_S^2 - 3) \over 12}\mathbb{S}_i + {k_S^2 (k_S^2 - 3) \over 36} \partial_i \mathbb{S} + {k_S^2 (k_S^2 - 3 \Omega^2) \over 24} \hat \nabla^j \mathbb{S}_{ji} \right]  \cr
  &= e^{-i\Omega\tau/L} \left[
   {\Omega^2 k_S (k_S^2-3) \over 12} -
     {k_S^3 (k_S^2-3) \over 36} +
     {k_S (k_S^2-3\Omega^2) (k_S^2-3) \over 36}
   \right] \mathbb{S}_i
 }
To get to the last line (which obviously vanishes), one uses the first relation in \eno{SRelations} on the second term of the previous line and the third relation in \eno{SRelations} on the third term.

\subsection{Approximate formulas for the quasinormal frequencies}
\label{FITS}

We find that all the quasinormal frequencies for the main series of the tensor, vector, and scalar modes are approximately described by
 \eqn{FitMain}{
  \Omega &= \Omega_1 + \Delta \Omega (q-1) \qquad q \geq 1\cr
  \Omega_1/\rho_H &= (3.195\pm 0.009)-(2.743\pm 0.001) i\cr
  \Delta \Omega/\rho_H &= (2.020\pm 0.003) - (2.0061\pm 0.0005) i\,
 }
for $\rho_H = 6$, $13$, and $20$, and $3 \leq n \leq 6$.  (Note that we are still setting $L=1$.)  Here, the uncertainties in the regression coefficients are simply the standard errors that we obtain from a simultaneous fit of all the frequencies from tables~\ref{TensorFreqTable}, \ref{VectorFreqTable}, and~\ref{ScalarFreqTable}.  Inspection of these tables shows some trends not captured in \eno{FitMain}, for example weak dependences on $n$ and $\rho_H$.\footnote{An example of such a trend in the scalar modes is the slight increase in the magnitudes of $\Re\Omega_0$ and $\rho_H \Im\Omega_0$ as $\rho_H$ increases.  An apparent exception is the real part for the $n=3$ case.  Closer inspection of the Wronskian function in this case shows that the numerics are not as reliable as usual for $\rho_H/L = 20$.}

The temperature as measured with respect to global time is $T_g \approx 2\rho_H / \pi L^2$ for $\rho_H \gg L$.  See \eno{GAdSBHmts} for a form that is valid for arbitrary $\rho_H/L$.  It is striking that $\Delta\Omega/2\pi T_g L \approx 1-i$.  If we revert to the notation $\omega = \Omega/L$, then the imaginary part of $\Delta\omega$ is approximately the first Matsubara frequency.  This clearly calls for some interpretation in terms of a Euclidean continuation.

The low-lying \emph{vector} modes are approximately given by
 \eqn{FitVector}{
  \Omega = -{i a\over 4 \rho_H} (n^2 - n_0^2) \qquad a = 1.018 \pm 0.008  \qquad n_0^2 = 4.14 \pm 0.17
 }
for $\rho_H = 6$, $13$, and $20$, and $3 \leq n \leq 6$.  For the same range of $n$ and $\rho_H$ we find that the quasinormal frequencies of the low-lying \emph{scalar} modes are given by
 \eqn{FitScalar}{
  \Omega &= {a_r \over \sqrt{3}} \sqrt{n^2 - n_{0, r}^2} - {i a_i \over 6 \rho_H} (n^2 - n_{0, i}^2) \cr
  a_r &= 1.031 \pm 0.012 \qquad n_{0, r}^2 = 1.46 \pm 0.41\cr
  a_i &= 0.985 \pm 0.007 \qquad n_{0, i}^2 = 3.91 \pm 0.14\,.
 }
An independent calculation of the low-lying vector and scalar modes can be done using linearized hydrodynamics on $S^3 \times \mathbf{R}$, which we will do in section \ref{HYDRO}.  Formulas \eqref{FitVector} and \eqref{FitScalar} are written in a form that makes easy the comparison to predictions from linearized hydrodynamics, \eqref{VectorRootAgain} and \eqref{ScalarRootsAgain}.

\begin{figure}
  \centerline{\includegraphics[width=3in]{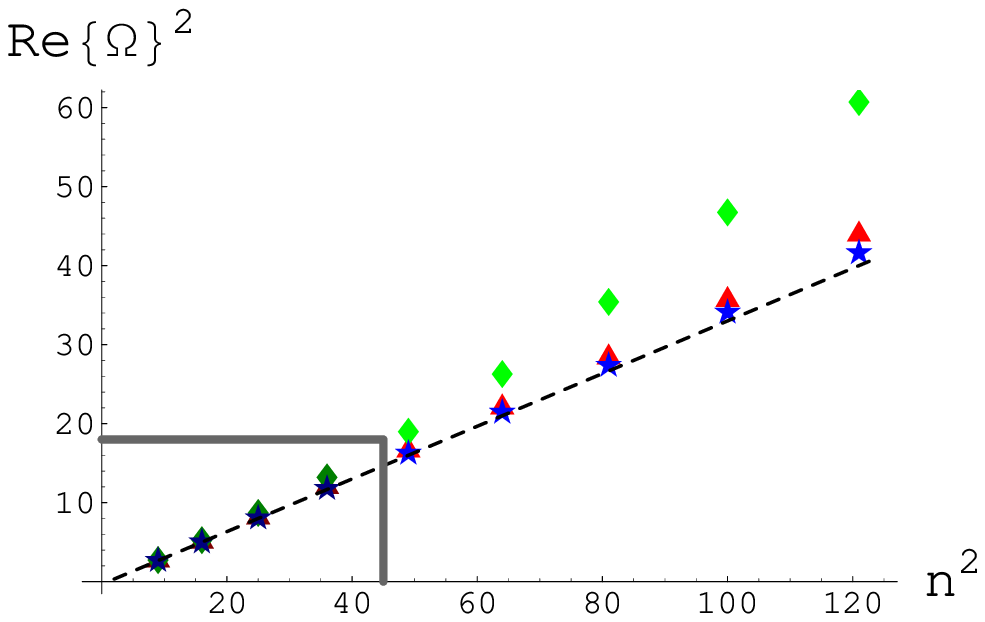}
  \hspace{0.25in}
  \includegraphics[width=3in]{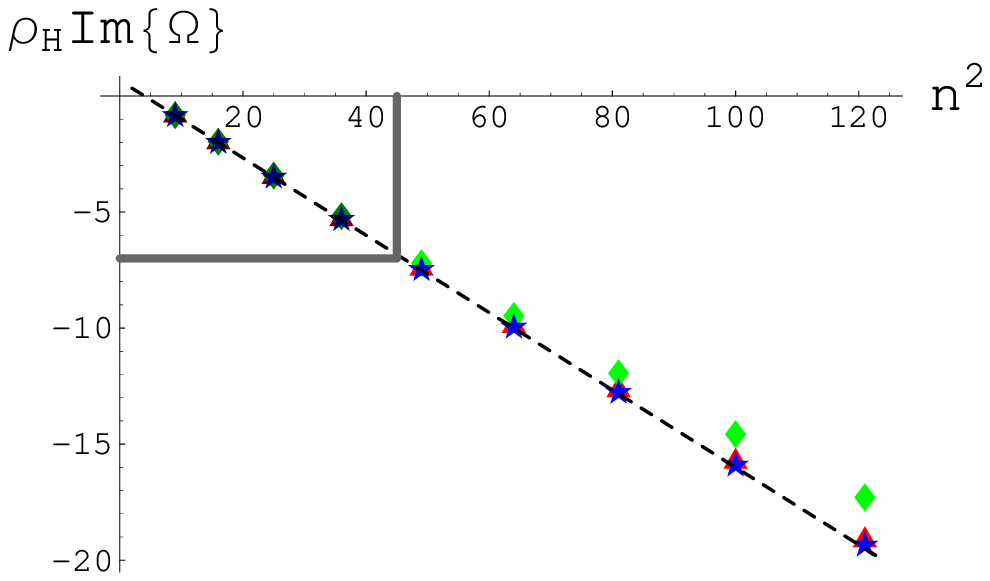}}
  \caption{The real (left) and imaginary parts (right) of the low-lying quasinormal frequencies for the scalar modes with $3 \leq n \leq 11$ (we have set $L=1$).  The green diamond-shaped points correspond to $\rho_H = 6$, the red triangular ones to $\rho_H = 13$, and the blue star-shaped ones to $\rho_H = 20$.  The dashed line shows the predictions from linearized hydrodynamics, as explained in section~\ref{HYDRO}.  The fit \eno{FitScalar} was done for the leftmost four points shown ($3 \leq n \leq 6$), but simultaneously for all three values of $\rho_H$---so $12$ data points altogether.  These points are the ones inside the inset frame in each plot.  The fit \eno{FitScalarAgain} comes from only the $\rho_H=20$ points.}\label{LowScalar}
 \end{figure}

We extended our calculations of the frequencies of scalar QNM's to higher $n$ and found that deviations from \eno{FitScalar} are fairly significant, especially for $\rho_H = 6$: see figure~\ref{LowScalar}.  On the other hand, for $\rho_H=20$, the fitting function in \eno{FitScalar} works well for $3 \leq n \leq 11$, but the best fit parameters change a little:
 \eqn{FitScalarAgain}{
  a_r &= 1.0170 \pm 0.0021 \qquad n_{0, r}^2 = 1.41 \pm 0.16\cr
  a_i &= 0.9922 \pm 0.0006 \qquad n_{0, i}^2 = 3.86 \pm 0.04\,.
 }
As is evident from figure~\ref{LowScalar}, the $\rho_H=13$ frequencies are fairly close to the values for $\rho_H=20$.

The overall pattern of quasinormal frequencies is similar to that found in \cite{Kovtun:2005ev}.  As noted earlier, this is not accidental: large $\rho_H/L$ means that replacing $S^3$ by ${\bf R}^3$ is for many purposes a good approximation.

\subsection{Summary of results}
\label{SUMMARY}

Let us summarize in a single formula all that we have learned about the stress tensor profiles dual to perturbed GAdSBH's:
 \eqn{ExpandT}{
  \langle T_{mn} \rangle =
   \langle T_{mn}^{\rm Sch} \rangle +
    {1 \over W^2} \sum_R Q_R \; (\tau/L)^{p(R)} \; e^{-i\Omega_R \tau/L} \;
      \eta_{mn}^R \,,
 }
where $Q_R$ are complex constants.  The rest of the notation in \eno{ExpandT}
is explained in the next paragraph.  The expression \eno{ExpandT} is a large family of {\it exact} solutions to the equations
 \eqn{ConserveTraceless}{
  \nabla^m \langle T_{mn} \rangle = 0 \qquad
    \langle T^m{}_m \rangle = 0 \,.
 }
These solutions are dual to {\it approximate} descriptions of deformations of GAdSBH---approximate because we used the linearized Einstein equations.  It is worth reiterating a point from the end of section~\ref{TENSORSTRESS}: after plugging in formulas for the spherical harmonics developed in appendix~\ref{TOMITA} and using \eno{Wdef} and \eno{FindTauChi}, the components of $\langle T_{mn} \rangle$ are explicit algebraic expressions in terms of $t$, $r$, and trigonometric functions of $\theta$ and $\phi$.\footnote{An exception is the case   where some $p(R)$ is non-vanishing, in which case powers of logs of rational expressions in $t$ and $r$ would also be involved.  However, as noted in the main text, non-vanishing $p(R)$ is only a theoretical possibility, and we did not encounter any examples in actual computations.}

In \eno{ExpandT}, $\langle T_{mn}^{\rm Sch} \rangle$ is the unperturbed conformal soliton flow \eno{FoundBackgroundT}, $N^2$ is the dimension of the gauge group, $L$ can be regarded as an adjustable parameter (because of dilation symmetry), and $W$ is the conformal factor defined in \eno{Wdef}.  $R$ is a ``composite index'' that labels a particular basis vector in a particular representation of $O(4)$ and a quasinormal frequency appropriate to that representation.  For convenience we also include a choice of $\rho_H/L$ in the information that $R$ specifies, but actually this choice is fixed once we select the background flow $\langle T^{\rm Sch}_{mn} \rangle$.  In other words, $\rho_H/L$ is the same for all the terms in \eno{ExpandT}.  Consider for example
 \eqn{ExampleR}{
  R \ =\ \left\{ \hbox{tensor},\ \hbox{even},\ n=5,\ \ell=3,\
   m = -1,\ \rho_H/L = 20,\ \Omega \approx 66.4-53.9i,\
   p=0 \right\} \,.
 }
The first three elements of this list select the $O(4)$ representation as the symmetric part of $7_L \otimes 3_R \;\oplus\; 3_L \otimes 7_R$ (see the discussion around \eno{RepContent}).  The next two elements of the list select the basis vector corresponding to $\mathbb{T}_{ij}^{\rm even}(5,3,-1)$.  (See \eno{TnlmEven} for an explicit expression for $\mathbb{T}_{ij}^{\rm even}(5,3,-1)$.)  To form $\eta_{mn}^R$ we refer to \eno{etaTensor}.  For vector or scalar modes we would refer instead to \eno{etaVector} or~\eno{etaScalar}.

Once the representation and $\rho_H/L$ are selected, there are discretely many possible quasinormal frequencies, namely the roots of $\psi(1,\Omega)$ where $\psi(y,\Omega)$ is a solution of the modified master equation \eno{psiEOM} (note however that a more sophisticated prescription is needed in the case of scalar QNM's, as described in section~\ref{SCALAR}).  In sections~\ref{TENSORFREQS}, \ref{VECTORFREQS}, and~\ref{SCALARFREQS} we made a numerical determination of some of these frequencies, as summarized in tables~\ref{TensorFreqTable}, \ref{VectorFreqTable}, and~\ref{ScalarFreqTable} and the approximate fitting forms discussed in section~\ref{FITS}.  If a particular root of $\psi(1,\Omega)$ has some multiplicity $J>1$, then the allowed time dependences at this frequency are $\tau^p e^{-i\Omega\tau/L}$ where $0 \leq p \leq J-1$.  These time dependences can be thought of as arising from differences of pure exponentials with infinitesimally separated frequencies $\Omega$.  A familiar example is the critically damped simple harmonic oscillator.  Although we have included in \eno{ExpandT} the possibility of non-zero $p$, our numerical explorations yielded only single roots of $\psi(1,\Omega)$.

The case of vector perturbations in the adjoint of $O(4)$ ($n=2$), as remarked around \eno{FieldFab}, is special in that there is only one mode, and it has $\Omega=0$ and $\eta_{mn}$ given by \eno{etaZeroMode}.

The only input from black hole physics (at the level of the current discussion) is the particular values of $\Omega$ that we must use.  The spectrum of possible $\Omega$ is symmetric under $\Omega \to -\Omega^*$, so by appropriately choosing the complex coefficients $Q_R$ one may ensure that $\langle T_{mn} \rangle$ is real.  Although $\langle T_{mn} \rangle$ is {\it exactly} traceless and conserved for any choice of the $Q_R$ (and indeed for arbitrary $\Omega$), in general it might have pathologies like negative energy density at sufficiently early times (e.g.~before the ``collision'').  Thus we should regard the perturbed conformal soliton as a framework best justified for addressing the late-time behavior of colliding conformal matter.  To achieve a better description of early times via AdS/CFT, it would be better to avoid the linearized approximation to Einstein's equations and consider some physical collision in anti-de Sitter space which results in a black hole that ``rings down'' through QNM's at late times towards GAdSBH or some other stationary solution.  For example, one could consider colliding shock waves, as advocated as a description of RHIC collisions in \cite{Nastase:2005rp}, or colliding black holes.

\section{Case studies of perturbed conformal soliton flows}
\label{CASES}

The aim of this section is to take the extended technical treatment of quasinormal modes of the GAdSBH background \eno{GAdSBHmetric} from section~\ref{SUMMARY} and attempt to extract some flows in the boundary gauge theory that more closely resemble heavy-ion collisions than the original conformal soliton \eno{FoundBackgroundT}.  There is a problem with this program at the level we are able to approach it in this paper: it's not clear {\it a priori} that the late-time regime in which QNM's are a good description of the black hole dynamics overlaps with the early time regime in which the temperature is high enough for conformal invariance to be an approximate symmetry of the QGP.  The situation might be improved somewhat by going to higher temperatures on the experimental side, as LHC experiments will do.  On the theoretical side, it is possible in principle to go beyond the QNM approximation with numerical simulations of the full Einstein equations.  It is worth recalling that although a real-world QGP at RHIC would be fully hadronized by times $t \gg L/c = 7\,{\rm fm}/c$, a truly conformal plasma (such as a ${\cal N}=4$ super-Yang-Mills plasma) would simply continue to cool and expand without going through any phase transitions.

Given the limitations just described, it is best to think of the studies below as preliminary forays intended to help define the questions we might ask of more substantial explorations.

\subsection{Slow vector modes}
\label{SLOWVECTOR}

\begin{figure}
  \centerline{\includegraphics[width=5in]{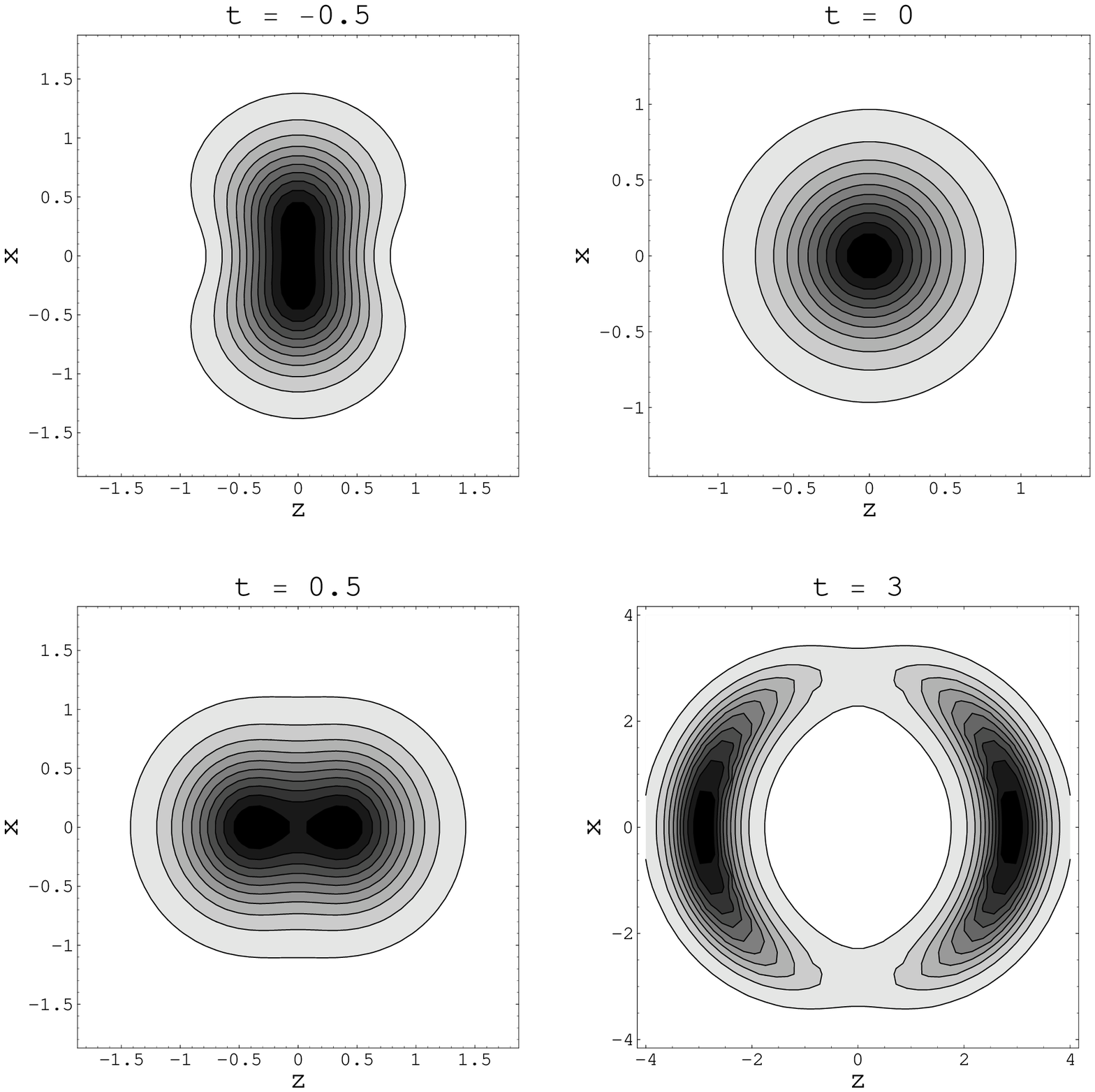}}
  \caption{Snapshots of the energy density of the flow specified by \eno{ExpandEps} with $Q_1 = 8 \times 10^4$.  The $z$ coordinate measures position along the beamline (where $\theta = 0$ or $\pi$), while the $x$ coordinate is transverse.  The flow is azimuthally symmetric, which is to say symmetric around the $z$ axis.  Note that the scales of the $x$ and $z$ axes change from frame to frame.}\label{SlowFlow}
 \end{figure}

Consider the low-lying vector modes whose quasinormal frequencies are given approximately in \eno{LowQNMs}, together with the zero modes described in the subsequent paragraph.  Simply for the purpose of getting started, let us consider the mode specified by
 \eqn{ExampleRone}{
  R_1 \ =\ \left\{ \hbox{vector},\ \hbox{even},\ n=3,\ \ell=2,\
   m = 0,\ \rho_H/L = 20,\ \Omega \approx -{i \over 16},\
   p=0 \right\} \,.
 }
From now on, we will mostly set $L=1$, which is equivalent to measuring all lengths in units of $L = 7\,{\rm fm}$.  The energy density and the radial Poynting vector are
 \eqn{ExpandEps}{
  \epsilon &= \langle T_{00} \rangle = \epsilon^{\rm Sch}(t,r) +
    Q_1 \; \epsilon_{R_1}(t,r,\theta,\phi) + \ldots  \cr
  S_r &= -\langle T_{0r} \rangle = S^{\rm Sch}_r(t,r) +
    Q_1 \; S_{r,R_1}(t,r,\theta,\phi) + \ldots \,.
 }
Choosing $Q_1 = 8 \times 10^4$ and $N^2=8$ results in a flow illustrated in figure~\ref{SlowFlow}.\footnote{Choosing $Q_1 = 10^4 N^2$ for any $N$ results in precisely the same flow up to an overall normalization.}  This flow is reminiscent of a Lorentz-flattened pancake expanding in something resembling Bjorken flow.  Alternatively, if the $z$ axis is replaced by the $y$ axis so that the views in figure~\ref{SlowFlow} are along the beamline, the flow appears somewhat similar to elliptic flow, where an initially anisotropic state expands faster in along its shorter axis.

The Bjorken flow analogy of the previous paragraph might contain a grain of truth, but we believe that the physics of the solution \eno{ExpandEps} is wrong for describing elliptic flow.  Consider the following points regarding the solution \eno{ExpandEps}:
 \begin{enumerate}
  \item The energy density is spherically symmetric at $t=0$.  This feature is common to all flows constructed by deforming the radial conformal soliton only by vector and tensor perturbations.
  \item The radial motion is, on average, inward rather than outward between $t=-0.5$ and $t=0$.  After $t=0$ the radial motion is mostly outward.  (Recall that we have set $L=1$.  If we restored dimensions with $L=7\,{\rm fm}$, $t=-0.5$ would translate into $t=-3.5\,{\rm fm}/c$.)
  \item There is negative energy density for times earlier than approximately $t=-0.56$.  If $Q_1$ were increased, there would be negative energy up to a later time.  Negative energy density at sufficiently early times is common to most linear perturbations \eno{ExpandT} of the conformal soliton flow, because the perturbations are signed quantities that grow exponentially as $t \to -\infty$.  This pathology is presumably cured when one goes beyond linear perturbation theory in the GAdSBH background.
  \item Referring to the $t=-0.5$ snapshot, one sees that the energy density profile is flattened in the $z$ direction by at best a factor of a few.  Real gold-gold collisions start with an aspect ratio possibly as high as $100$, although by the earliest time that thermalization is expected, this aspect ratio is probably closer to $10$.
 \end{enumerate}
From the earliest time ($t \approx -0.56$) that the linear approximate can be used without encountering negative energy density, the radial component of the Poynting vector is more positive on the horizontal axis of figure~\ref{SlowFlow} than on the vertical axis.  Moreover, $\Omega$ is purely imaginary.  So instead of describing some compressed state accelerating preferentially along this horizontal axis, it seems that we have described a flow that starts with outward velocities preferentially horizontal and that manages to keep expanding horizontally despite some modest damping.  This is not how elliptic flow is thought to arise.

\subsection{Slow scalar modes and elliptic flow}
\label{SLOWSCALAR}

\begin{figure}
  \centerline{\includegraphics[width=5in]{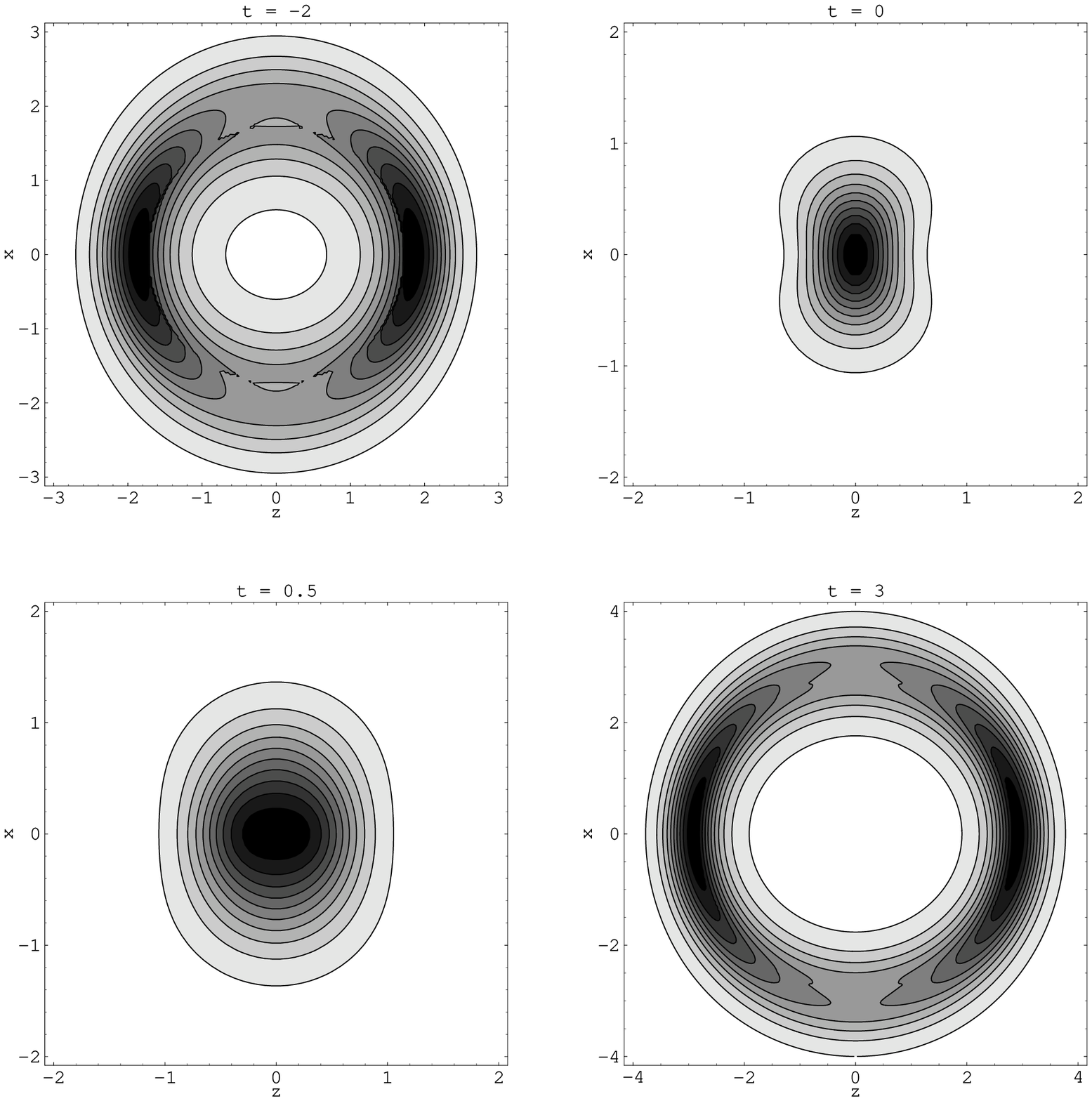}}
  \caption{Snapshots of the energy density of the flow specified by \eno{ExpandScalarEps} with $Q_2 = -10^4$.  The $z$ coordinate measures position along the beamline (where $\theta = 0$ or $\pi$), while the $x$ coordinate is transverse.  The flow is azimuthally symmetric, which is to say symmetric around the $z$ axis.  Note that the scales of the $x$ and $z$ axes change from frame to frame.}\label{ScalarFlow}
 \end{figure}

Now consider the low-lying scalar modes whose quasinormal frequencies are given approximately in \eno{FitScalar}.  To get started, let us consider the mode specified by
 \eqn{ExampleRtwo}{
  R_2 \ =\ \left\{ \hbox{scalar},\ n=3,\ \ell=2,\
   m = 0,\ \rho_H/L = 13,\ \Omega \approx 1.637 - 0.064i,\
   p=0 \right\} \,.
 }
We will again set $L=1$ in the rest of the discussion.  In analogy to \eno{ExpandEps}, we exhibit an explicit special case of \eno{ExpandT} as
 \eqn{ExpandScalarEps}{
  \epsilon &= \langle T_{00} \rangle = \epsilon^{\rm Sch}(t,r) +
    2 \Re\left\{ Q_2 \; \epsilon_{R_2}(t,r,\theta,\phi)
       \right\} + \ldots  \cr
  S_r &= -\langle T_{0r} \rangle = S^{\rm Sch}_r(t,r) +
    2 \Re\left\{ Q_2 \; S_{r,R_2}(t,r,\theta,\phi)
       \right\} + \ldots \,.
 }
We must explicitly take the real parts in \eno{ExpandScalarEps} because the quasinormal frequencies are complex.  This is in contrast to \eno{ExpandEps}, where the frequencies are purely imaginary and the quasinormal mode expressions are already real.  Choosing $Q_2 = -10^4$ and $N^2=8$ results in a flow illustrated in figure~\ref{ScalarFlow}.  (As before, the same flow up to a normalization would result from choosing any $Q_2$ and $N^2$ with $Q_2/N^2 = -10^4/8$.)
Once again, we can attempt an analogy to the longitudinal collective motion of a central collision, or, by changing axes $z \to y$ so that the plots in figure~\ref{ScalarFlow} are views down the beampipe, elliptic flow in a non-central collision.  Both analogies may have some merit, but let us focus on the possible connection to elliptic flow.  At $t=0$, the radial and tangential components of the Poynting vector are small: typically on the order of $10\%$ or less of the change $|Q_2 \epsilon_{R_2}|$ in the energy density.  This is in stark contrast with the situation for vector modes, where $\epsilon_{R_1}=0$ at $t=0$ while the Poynting vector is non-zero.  The smallness of the Poynting vector means that we are starting at $t=0$ with an approximately stationary plasma that has been compressed into an elliptical shape.  Its preferential expansion along the short axis of the ellipse must then be due to the pressure gradients.  This reasoning is reinforced by the observation that $\Omega$ is almost real: the real part describes the ``springiness'' of the plasma, whereas the small imaginary part describes damping.

We can go further and obtain an expression for the elliptic flow coefficient $v_2$ normalized by the eccentricity of the initial state at $t=0$.  To this end, let us explicitly consider the $SO(3)$ rotation that rotates the $z$ axis into the $y$ axis.  After this rotation, there are two more modes involved:
 \eqn{ExampleRthreefour}{
  R_3 \ &=\ \left\{ \hbox{scalar},\ n=3,\ \ell=2,\
   m = 2,\ \rho_H/L = 13,\ \Omega \approx 1.637 - 0.064i,\
   p=0 \right\}  \cr
  R_4 \ &=\ \left\{ \hbox{scalar},\ n=3,\ \ell=2,\
   m = -2,\ \rho_H/L = 13,\ \Omega \approx 1.637 - 0.064i,\
   p=0 \right\} \,.
 }
Consider an expansion
 \eqn[c]{DwaveExpand}{
  \epsilon = \langle T_{00} \rangle = \epsilon_0 + \epsilon_1 +
    \ldots \qquad\qquad
  \epsilon_0 \equiv \epsilon^{\rm Sch}(t,r)  \cr
  \epsilon_1 \equiv
    2 \Re\left\{ \tilde{Q}_2 \; \epsilon_{R_2}(t,r,\theta,\phi) +
     \tilde{Q}_3 \; \epsilon_{R_3}(t,r,\theta,\phi) +
     \tilde{Q}_4 \; \epsilon_{R_4}(t,r,\theta,\phi)
       \right\} \,,
 }
where $\ldots$ indicates other deformations that we do not include in subsequent calculations.  The coefficients corresponding to a $90^\circ$ rotation of \eno{ExpandScalarEps} are
 \eqn{rotatedQs}{
  \tilde{Q}_3 = \tilde{Q}_4 = -\sqrt{3 \over 8} Q_2
    \qquad
   \tilde{Q}_2 = -{1 \over 2} Q_2 \,.
 }
The eccentricity at time $t=0$ is
 \eqn{EccentricityDef}{
  \delta \equiv {\langle y^2-x^2 \rangle \over \langle y^2+x^2 \rangle}
     = -{\langle r^2 \sin^2\theta \cos 2\phi \rangle \over
           \langle r^2 \sin^2\theta \rangle}
     \approx {\int_{{\bf R}^3} \epsilon_1 \, r^2 \sin^2\theta \cos2\phi
       \over \int_{{\bf R}^3} \epsilon_0 \, r^2 \sin^2\theta} \,,
 }
The averages are computed with respect to a probability measure proportional to the energy density $\epsilon$ given in \eno{DwaveExpand}, evaluated at $t=0$.  The final expression in \eno{EccentricityDef} is approximate because we used $\epsilon_0$ in the denominator instead of $\epsilon = \epsilon=\epsilon_0+\epsilon_1$.

We will define $v_2$ to be $\langle \cos 2\phi \rangle$ evaluated at $\theta=\pi/2$ (mid-rapidity), where the average is again computed with respect to energy density, but now evaluated in the limit of late times.  The energy density $\epsilon$ is known in closed form, and it turns out that the integrals that define the averages can be done.  The result is
 \eqn{NormalizedVtwo}{
  {v_2 \over \delta} = {1 \over 6\pi} \Re\left\{
    {\Omega^4 - 40\Omega^2 + 72 \over \Omega^3 - 4\Omega}
    \sin {\pi\Omega \over 2} \right\} \,.
 }
Readers wishing to verify \eno{NormalizedVtwo} explicitly will be aided by the formula
 \eqn{epsilonRthree}{
  \epsilon_{R_3} &= -{8\sqrt{2/3} \over
     \pi \left[ r^4 - 2r^2 (-1+t^2) + (1+t^2)^2 \right]^4} e^{2i\phi}
     \left( r^2 - (i+t)^2 \over r^2 - (i-t)^2 \right)^{\Omega/2}
     r^2 \sin^2\theta  \cr &\qquad\qquad{} \times \Big[
       r^4 (-40-20i\Omega t - 8t^2 + 3\Omega^2 t^2) -
       2r^2 ( 40 + 64 t^2 + \Omega^2 t^2 - 8 t^4 + 3\Omega^2 t^4)
      \cr &\qquad\qquad\qquad{} +
       (1+t^2)^2 (-40 + 20i\Omega t - 8 t^2 + 3 \Omega^2 t^2) \Big]
    \,.
 }
The $\Omega$ corresponding to the low-lying $n=3$ scalar mode is not very sensitive to the choice of $\rho_H$: see table~\ref{ScalarFreqTable}.  Plugging in the value for $\Omega$ quoted in $R_2$, $R_3$, and $R_4$ gives $v_2/\delta = 0.37$.  This is at least in the right ballpark for comparison to experiment.  For example, in \cite{Adare:2006ti}, after $p_T$-averaging the elliptic flow coefficient over $0.3-2.5\,{\rm GeV}/c$ and estimating the eccentricity using a Glauber model, a value $\delta/v_2 \approx 3.1$ is found.

Weak points in our analysis include:
 \begin{enumerate}
  \item The formula \eno{NormalizedVtwo} appears highly predictive, but we should remember that there are $d$-wave excitations for higher $n$ as well.  Linear perturbation theory by itself cannot say whether the minimal $n=3$ $d$-wave mode is the dominant one.  This is one of many aspects where a better understanding of collisions of black holes in global $AdS_5$ (for example through numerical simulation) would be desirable.
  \item The configuration at $t=0$ is nearly spherical for small eccentricity $\delta$, whereas in heavy-ion collisions at RHIC, the $t=0$ configuration for small eccentricity is Lorentz-flattened into a disk.  Again, this is curable in principle if one could numerically solve the full Einstein equations for black hole collisions in $AdS_5$.
  \item We have neglected hadronization.  This could be fixed simply by appending a Cooper-Frye prescription \cite{Cooper:1974mv} to the evolution \eno{DwaveExpand} of the energy density.
 \end{enumerate}

\subsection{Fast modes and the thermalization time}
\label{FASTMODES}

Aside from the slow modes discussed in sections~\ref{SLOWVECTOR} and~\ref{SLOWSCALAR}, all deformations of the radial conformal soliton damp out quite quickly.  The time dependence of these modes is dominated by the factor $e^{-i\Omega\tau/L}$.  Referring to \eno{YtoX}, we see that $\tau \approx 2t$ near the origin, $t=r=0$.  So for $|t|$,~$|r| \lsim L$, the time-dependence of the fast modes is roughly $e^{-2i\Omega t/L}$, and we conclude that the timescale for their decay by a factor of $1/e$ is
 \eqn{tTherm}{
  \tau_{e-\rm fold} \sim {L \over 2 |\Im\Omega_1|} \approx
    {1 \over 8.6 \, T_{\rm peak}} \approx 0.08\,{\rm fm}/c \,.
 }
To obtain the first estimate in \eno{tTherm} we used the frequency $\Omega_1$ in the main sequence with the least negative imaginary part.  To obtain the second approximate equality we used \eno{TPeak}.  To obtain the third approximate equality we plugged $T_{\rm peak} = 300\,{\rm MeV}$, which is in the upper ranges of temperatures believed to be attained in thermalized states at RHIC.  It is natural to conjecture an extrapolation of the result \eno{tTherm} to more general conformations of ${\cal N}=4$ plasma: $1/\tau_{e-\rm fold} \approx 8.6 T_{\rm local}$ for the time $\tau_{e-\rm fold}$ that it takes deviations from local equilibrium to decay by $1/e$.

The thermalization time $\tau_{\rm therm}$ in RHIC collisions is believed to be in the vicinity of $0.6\,{\rm fm}/c$.\footnote{In \cite{Arnold:2004ti} the value $\tau_{\rm therm} \approx 0.6\,{\rm fm}/c$ is quoted, and in \cite{Adcox:2004mh} the range $0.6\,{\rm fm}/c \leq \tau_{\rm therm} \leq 1\,{\rm fm}/c$ is favored.  The basis for this range of values appears to be a survey of hydro calculations with various calculational methods as well as differing assumptions about the equation of state.}  Larger values tend to cause elliptic flow to be under-predicted by hydrodynamical simulations.  But it is a bit of a puzzle in QCD how such a fast equilibration time results.  Analysis of perturbative scattering processes seem to indicate $\tau_{\rm therm} \gsim 2.5\,{\rm fm}/c$ \cite{Baier:2000sb,Molnar:2001ux}.  The momentum scales of typical processes decrease quickly during the approach to thermal equilibrium in RHIC collisions, so the effective 't~Hooft coupling quickly increases away from the perturbative regime.  But it is not clear that the regime of couplings pertinent to equilibration is directly comparable to the $\lambda \to \infty$ limit of ${\cal N}=4$ SYM where supergravity gravity calculations become reliable.  So we should be more cautious than usual about attempting to apply \eno{tTherm} to RHIC collisions.  Nevertheless, it is comforting to see a short $e$-folding time in a strongly coupled plasma.  To put it another way: it would have been alarming to find $\tau_{e-\rm fold} > 1\,{\rm fm}/c$ from strongly coupled ${\cal N}=4$ SYM, where if anything one should expect faster thermalization than in real-world RHIC collisions.  Given the highly anisotropic momentum space distribution expected in the early stages of a RHIC collision, perhaps it is reasonable to expect that several $e$-folding times of the relevant thermalization / isotropization processes must elapse before hydrodynamic approximations can be used.  If so, then perhaps starting from \eno{tTherm} we should estimate $\tau_{\rm therm} \sim 0.3\,{\rm fm}/c$ for the analogous processes in ${\cal N}=4$ SYM.  One should bear in mind that this estimate incorporates considerable uncertainties.

One of the favored thermalization scenarios in RHIC collisions involves plasma instabilities (see for example the review talk \cite{Arnold:2004tf} and references therein), which are argued to be capable of producing a spatially isotropic form of the local stress tensor in a time comparable to $0.6\,{\rm fm}/c$ \cite{Arnold:2004ti}.  The plasma instabilities owe their existence to an anisotropic distribution of momenta and have their wave-vectors primarily along the beamline.  Perhaps some of the main sequence quasinormal modes describe the late stages of the evolution of analogs of these plasma instabilities.

\section{Discussion}
\label{DISCUSS}

Through the conformal transformation \eno{EmbedMinkowski} we have mapped a static plasma on $S^3 \times {\bf R}$ to a radial flow of thermal matter in Minkowski space which we have termed the ``conformal soliton.''  Our subsequent calculations of quasinormal modes in the GAdSBH background can be interpreted holographically as slight motions in the static plasma on $S^3$, or as perturbations to the radial flow.  A salient feature of these perturbations is that they oscillate and damp out exponentially in global time $\tau$, which runs over only a finite range of values across Minkowski space.  Indeed, the qualitative physics behind our estimate $v_2/\delta \sim 0.37$ is that a $d$-wave excitation has about enough global time to go through half an oscillation.

\subsection{Linearized hydrodynamics}
\label{HYDRO}

Given the separation by a factor of tens to hundreds (depending on the choice of $\rho_H/L$) between the imaginary parts of the low-lying scalar and vector modes and the ``main sequence'' modes, it is natural to regard the former as hydrodynamic and the latter as microscopic.  With this in mind, we might expect to reproduce the low-lying modes from linearized hydrodynamics.  It is easiest to think about linearized hydrodynamics on $S^3 \times {\bf R}$: the general equations are
 \eqn[c]{LinearizedHydro}{
  \tilde\nabla^a \tilde{T}_{ab} = 0  \cr
  \tilde{T}_{ab} = (\epsilon + p) u_a u_b + p \tilde{g}_{ab} +
    \tau_{ab}  \cr
  \tau_{ab} \equiv -\eta \left( \Delta_{ac} \tilde\nabla^c u_b +
    \Delta_{bc} \tilde\nabla^c u_a - {2 \over 3} \Delta_{ab}
    \tilde\nabla^c u_c \right) - \xi \Delta_{ab} \tilde\nabla^c u_c
      \cr
  \Delta_{ab} \equiv \tilde{g}_{ab} + u_a u_b \,,
 }
and the condition $\tilde{T}^a{}_a = 0$ implies both $\epsilon = 3p$ and $\xi=0$.  As usual, $u^a$ is a timelike vector with $u^2 = -1$.  The metric $\tilde{g}_{ab}$ is the one specified in \eno{NaturalSthreeMet}, and $\tilde\nabla_a$ is the associated connection.  A slight subtlety arises in defining $\tilde{T}_{ab}$ because of the presence of the Casimir energy, which makes a temperature-independent pure trace contribution to the stress tensor.  We will ignore this contribution and equate $\tilde{T}_{ab}$ with $\langle \tilde{T}_{ab} \rangle_{\rm sub}$ as appearing in \eno{EfficientT} and defined in the discussion preceding \eno{Wdef}, so that the unperturbed fluid has stress energy as given in \eno{FoundSthreeT}.

In the linearized approximation, $u_a = (-1,u_i)$ where $i$ runs over the $S^3$ directions and the $u_i$ are small.  One also must write a perturbed expression for the pressure, $p = p_0 + \delta p$, where $\delta p$ is the same order as the $u_i$.  The resulting equations are
    \eqn[c]{SimpLH}{
     3 \partial_\tau \delta p + 4 p_0 \tilde\nabla^i u_i = 0  \cr
     4 p_0 \partial_\tau u_i + \partial_i \delta p -
       \eta \left(\tilde\nabla^j \tilde\nabla_j u_i + \tilde\nabla^j \tilde\nabla^i u_j\right) +
       {2 \eta  \over 3} \partial_i (\tilde\nabla^j u_j) = 0 \,.
    }
To ease the notational burden, let us set $L=1$ for the rest of the discussion.

An appropriate ansatz to describe the low-lying scalar modes is
 \eqn{scalarAnsatz}{
  \delta p = K_1 e^{-i\Omega\tau} \mathbb{S} \qquad
   u_i = K_2 e^{-i\Omega\tau} \mathbb{S}_i \,,
 }
where $\mathbb{S}$ is a scalar harmonic on $S^3$ and $\mathbb{S}_i = -{1 \over k_S} \partial_i \mathbb{S}$ as in \eno{ScalarSDefs}.  Plugging \eno{scalarAnsatz} into \eno{SimpLH} leads to
    \eqn[c]{GotScalarOmega}{
     -3i\Omega K_1 + 4 p_0 k_S K_2 = 0  \cr
     -k_S K_1 + \left( -4 i p_0 \Omega +
       {4\eta \over 3} k_S^2 - 4\eta \right) K_2 = 0 \,.
    }
One may eliminate $K_1$ and $K_2$ to obtain a quadratic equation for $\Omega$.  For small $\eta$, the roots are
    \eqn{ScalarRoots}{
     \Omega = \pm {k_S \over \sqrt{3}} - {i(k_S^2-3) \over 6}
       {\eta \over p_0} + O(\eta^2) \,.
    }
To simplify \eno{ScalarRoots}, we use $p_0 = \epsilon_0/3$ and then combine the relations \eno{GAdSBHmts} for the total entropy $S$ and mass $M$ of the unperturbed GAdSBH solution into the form
 \eqn{etaPratio}{
  {\eta \over p_0} = {3\eta \over s} {S \over M} =
    {4\pi\eta \over s} {\rho_H \over 1+\rho_H^2} \,.
 }
To leading order in large $\rho_H$,
    \eqn{ScalarRootsAgain}{
     \Omega = \pm {k_S \over \sqrt{3}} -
        {i (k_S^2-3) \over 6 \rho_H} \,,
    }
where we have used \eno{etaPratio} and set $\eta/s = 1/4\pi$.  Equation~\eno{ScalarRootsAgain} is in good agreement with the results of numerical evaluation of low-lying scalar quasinormal frequencies for large $\rho_H$: see~\eno{FitScalar}, \eno{FitScalarAgain}, and figure~\ref{LowScalar}.

An analogous treatment can be made for the low-lying vector modes, where the appropriate fluid dynamics ansatz is
 \eqn{vectorAnsatz}{
  \delta p = 0 \qquad u_i = K_3 e^{-i\Omega\tau} \mathbb{V}_i \,.
 }
The first equation of (5.2) is satisfied trivially, and the second leads immediately to a linear equation for $\Omega$
     \eqn{GotVectorOmega}{
      -4i p_0 \Omega + \eta \left(k_V^2-2\right) = 0
     }
    whose root is
     \eqn{VectorRoot}{
      \Omega = -i{k_V^2 - 2 \over 4} {\eta \over p_0} \,.
     }
    Using (5.6) and $\eta/s = 1/4\pi$, one finds to leading order in large $\rho_H$ that
     \eqn{VectorRootAgain}{
      \Omega = -i{k_V^2-2 \over 4\rho_H} \,,
     }
    which is in good agreement with the fit (3.43) to our numerical evaluations of low-lying vector quasinormal frequencies.

A hydrodynamic ansatz cannot be constructed from the spherical harmonics $\mathbb{T}_{ij}$ and their derivatives because of the conservation and tracelessness properties of $\mathbb{T}_{ij}$.  This tallies with the absence of low-lying tensor quasinormal frequencies.

\subsection{Boosted conformal solitons}
\label{BOOSTED}

As we have previously remarked, a natural extension of the linearized quasinormal mode calculations presented in this paper would be a finite-element analysis of colliding black holes.  To gain some intuition for what such an analysis might show, we present the results of two small calculations: first, in this section, the holographic ``shadow'' $\langle T_{mn} \rangle$ of two black holes about to collide; and second, in section~\ref{TESTMASSES} the geometry in $AdS_5$ of massive test particles on near-collision trajectories.  In section~\ref{COLLIDE} we make some heuristic remarks about black hole collisions.

\begin{figure}
  \centerline{\includegraphics[width=6.5in]{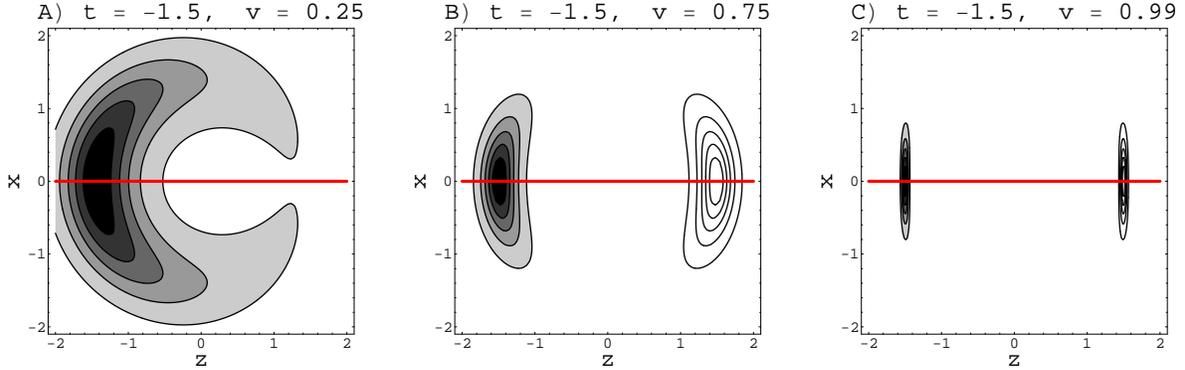}}
  \caption{Energy contours of boosted conformal solitons.  $L$ has been set to $1$.  The red line indicates the ``beam line,'' parallel to the center-of-mass momentum.  (A) shows that at slower velocities one starts to recover the picture of radial inflow.  In (B) and (C) we have indicated an additional boosted conformal soliton (unshaded contours) moving in the opposite direction, with zero impact parameter.  The energy density is azimuthally symmetric around the beam line.}\label{LorentzFlattened}
 \end{figure}

Recall that the conformal soliton describes a spherical shell of conformal matter that first implodes and then expands.  It is well understood that a Lorentz boost of a spherically symmetric shell expanding at the speed of light is still a spherical shell: only the energy density is no longer even distributed over the sphere, so that the center of mass moves.  This does not seem very like a gold nucleus moving at some speed $v$.  But as $v \to 1$, most of the stress-energy is concentrated near the ``beamline,'' and the boosted radial conformal soliton begins to resemble a Lorentz-flattened pancake, with transverse size roughly $2L$ and thickness roughly $2L/\gamma$.  See figure~\ref{LorentzFlattened}.

\subsection{Test masses in $AdS_5$}
\label{TESTMASSES}

It is interesting to recall that the conformal soliton depends on the size $\rho_H$ of the black hole only through a normalization.  So {\it exactly} the same contour plots as shown in figure~\ref{LorentzFlattened} would result from considering the holographic shadow of very large or very small black holes moving through $AdS_5$, only with a different scale for the energy density.  In the small $\rho_H$ limit we may replace the black holes with test masses moving on timelike geodesics in $AdS_5$.  All such geodesics are $SO(4,2)$ transformations of the trivial one, $\rho=0$, representing a test mass at rest (and hence an unboosted conformal soliton with very small amplitude).  Let us trace through the steps to describe the test mass trajectories corresponding to a head-on collision with each mass moving at speed $v$:
 \begin{enumerate}
  \item In Minkowski space, the trajectory of the center of mass of the right-moving soliton is described by the equations
 \eqn{xto}{
  x^3 = v t \qquad x^1 = x^2 = 0 \,.
 }
  \item Using \eno{globalToX} and our knowledge of the special case $v=0$, we reason that the appropriate trajectory of the test mass in $AdS_5$ is described by
 \eqn{Xto}{
  X^3 = v X^0 \qquad X^1 = X^2 = X^4 = 0 \,.
 }
This trajectory, like any timelike geodesic in $AdS_5$, is the intersection of the hyperboloid \eno{Xmanifold} with a two-dimensional plane (in this case \eno{Xto}) that passes through the origin.
  \item Using \eno{globalToY}, one obtains from \eno{Xto} the following description of the test mass trajectories:
 \eqn{yto}{
  {\rho/L \over \sqrt{1+\rho^2/L^2}} = \left| v\sin {\tau \over L}
    \right| \qquad
   \chi = {\pi \over 2} \qquad \theta = 0\ \ \hbox{or} \ \ \pi \,.
 }
At $\tau=0$, when the test mass corresponding to the right-moving conformal soliton passes through $\rho=0$, its $\theta$ switches from $\pi$ to $0$.  The opposite happens for the left-moving test mass.  Note that there is only one intersection point of the two geodesics in the wedge of $AdS_5$ covered by Poincar\'e coordinates (see figure~\ref{GlobalBH}a).
  \item Using \eno{globalToX} directly, or \eno{globalToPoincare} and \eno{yto}, one may show that the trajectory of the right-moving test mass in the Poincar\'e patch is described by \eno{xto} together with
 \eqn{zGeodesic}{
  {z \over L} = \sqrt{1 + (1-v^2)t^2/L^2} \,.
 }
 \end{enumerate}
Now consider a non-central collision of boosted conformal solitons.  We start by offsetting each one in the $x$ direction by $\pm b/2$, where $b$ is the impact parameter.  This is trivially implemented in the Poincar\'e patch description: the trajectory of the right-moving test mass is
 \eqn{RightMovingMass}{
  x^3 = v t \qquad x^1 = {b \over 2} \qquad x^2 = 0 \qquad
   {z \over L} = \sqrt{1 + (1-v^2)t^2/L^2} \,,
 }
and for the left-moving mass one reverses the signs of $v$ and $b$.  Using~\eno{globalToX} one immediately finds the description of the trajectory \eno{RightMovingMass} in terms of the hyperboloid \eno{Xmanifold} cut by the plane
 \eqn{CuttingPlane}{
  X^3 = v X^0 \qquad X^1 = {b \over 2L} (X^{-1} + X^4) \qquad
    X^2 = 0 \qquad X^4 = -{b \over 4L} X^1 \,.
 }
Using \eno{globalToY} one may rewrite \eno{CuttingPlane} (after some algebra) as
 \eqn{RightMoverOffCenter}{
  {\rho/L \over \sqrt{1+\rho^2/L^2}} =
    {v \sin {\tau \over L} \over \sin\chi \cos\theta} \qquad
   \tan\chi = -{4L \over b\sin\theta} \qquad
   \tan\theta = {4b \over vL} {\cot {\tau \over L} \over
     8+b^2/L^2} \,.
 }
It is straightforward to solve \eno{RightMoverOffCenter} explicitly for $\rho$, $\chi$, and $\theta$ in terms of $\tau$, but the resulting expressions are long and not very enlightening.

\subsection{Colliding black holes in $AdS_5$}
\label{COLLIDE}

Black hole collisions are understood to be highly inelastic, with a large fraction of the total energy going into a single final state black hole.  The collision of black holes much smaller than the $AdS_5$ radius ($\rho_H / L \ll 1$) could be described approximately as a collision in flat space.  The four-dimensional case is a classic problem of numerical general relativity, and considerable progress has been made on it recently: see for example \cite{Pretorius:2005gq,Pretorius:2006tp}.  To make the closest possible analogy with RHIC collisions, one should consider highly relativistic collisions of black holes with $\rho_H \gg 1$ in five dimensions---a significantly harder problem!  But unlike in the astrophysically relevant case, one does not need to evolve for a long time: global time need only run from $\tau \approx -\pi L$ to $\pi L$.  The remarkable separation in Minkowski space of conformal solitons boosted in opposite directions, illustrated in figure~\ref{LorentzFlattened}, suggests that when the speed $v$ is close to $1$, even large black holes are in some sense well-separated and non-interacting in $AdS_5$ until a global time $|\tau| \ll L$, possibly as small as $2L/\gamma$.  However, it is understood at least in the flat space case that a horizon forms around colliding shock waves that encloses a large percentage of the energy \cite{Penrose74,DEath:1992hb,DEath:1992hd,DEath:1992qu,Eardley:2002re}.  For purposes of describing the subsequent ``ring-down,'' the geometry outside the newly formed horizon is more relevant than the microscopic interactions inside it because the latter are, by definition, causally inaccessible to an outside observer.  A thought-provoking idea is that some holographic translation of this picture should be relevant to heavy-ion collisions.  The possible relation of horizon formation from colliding shock waves in $AdS_5$ to RHIC physics has been explored already in \cite{Nastase:2005rp,Shuryak:2005ia,Nastase:2006eb}.  Heuristically, horizon formation means that perturbative processes early in the collision are washed out, or masked, by subsequent strongly interacting dynamics.  For certain high $p_T$ processes, such as the production of direct photons with $p_T \gsim 4\,{\rm GeV}/c$, this can hardly be true.  Nevertheless, it seems to us plausible that horizon formation in $AdS_5$ might have to do with opacity properties of the QGP.  Clearly it would desirable to supplement these highly speculative notions with calculations relevant to RHIC observables.

\subsection{Summary}
\label{CONCLUSIONS}

We have shown how to describe an expanding plasma of finite extent in terms of the holographic image of a global $AdS_5$ black hole (GAdSBH).  The key step is to use a conformal transformation to map a patch of $S^3 \times {\bf R}$ (the boundary of global $AdS_5$) onto ${\bf R}^{3,1}$.  Before this transformation, the unperturbed black hole is dual to a plasma at rest on $S^3$.  After the transformation, the unperturbed plasma flow is entirely radial, first contracting and then expanding.  In order to describe a plasma whose extent (at the time of maximal compression) is much greater than its reciprocal temperature, we must consider black holes whose horizon radius $\rho_H$ is much larger than the scale of curvature $L$ of $AdS_5$.

The bulk of our calculations are, as far as we know, the first full treatment of gravitational quasinormal modes of the five-dimensional GAdSBH background.  These perturbations are dual to slightly anisotropic expansions of the plasma.  For large black holes, there is a clean separation of modes into hydrodynamic and microscopic modes.  The microscopic modes decay with an $e$-folding time that is comparable to the reciprocal of the first Matsubara frequency, suggesting a thermalization time $\tau_{\rm therm} \sim 0.3\,{\rm fm}/c$ for temperatures and initial anisotropies comparable to those at RHIC collisions.  Analysis of one of the hydrodynamic modes leads to an estimate $v_2/\delta \approx 0.37$, where $v_2$ is the elliptic flow coefficient and $\delta$ is the eccentricity of the initial state.

The reader is emphatically warned that applications of our calculations to heavy ion collisions rest on assumptions and uncontrolled approximations.  Among them:
 \begin{enumerate}
  \item We have replaced QCD by ${\cal N}=4$ super-Yang-Mills theory in the limit of strong coupling.
  \item We rely particularly heavily on conformal invariance in passing between $S^3 \times {\bf R}$ and Minkowski space.
  \item The unperturbed motion of the plasma is radial rather than longitudinal.\label{ItsRadial}
 \end{enumerate}
We have already remarked that point~\ref{ItsRadial} could perhaps be ameliorated by better analyzing collisions of black holes or shock waves in $AdS_5$.  Our quasinormal mode analysis was entirely based on linearizing Einstein's equations around a globally static background, and it does not seem likely that QNM's are wholly adequate to describe the evolution of a horizon that forms from a high-speed collision.  However, the qualitative successes of the simplest case studies in section~\ref{CASES} suggest that QNM's will continue to play a role even once more sophisticated analyses or full-fledged numerical studies become available.

\section*{Acknowledgments}

We thank P.~Goldreich, J.~Maldacena and L.~Yaffe for helpful discussions.  This work was supported in part by the Department of Energy under Grant No.\ DE-FG02-91ER40671, and by the Sloan Foundation.

\clearpage
\appendix
\section{Spherical harmonics on $S^3$}
\label{TOMITA}

Scalar, vector, and tensor spherical harmonics on $S^3$ are well studied, in part because of their utility in describing perturbations of the closed four-dimensional FRW universe \cite{Lifshitz:1963ps}.  Other early treatises include \cite{Jantzen78,Sandberg82,Tomita:1982ew}, and a more general account of harmonics on $S^N$ for conserved, symmetric, traceless tensors of rank $r$ was given in \cite{Higuchi:1986wu}.  Mostly we follow \cite{Tomita:1982ew}.

We follow standard notation for scalar spherical harmonics on $S^2$:
 \eqn{YlmDef}{
  Y_{\ell m}(\theta,\phi) =
    \sqrt{{2\ell+1 \over 4\pi} {(\ell-m)! \over (\ell+m)!}} \,
     P_\ell^m(\cos\theta) e^{im\phi} \,,
 }
where the associated Legendre polynomials are
 \eqn{LegendreP}{
  P_\ell^m(x) = {(-1)^m \over 2^\ell \ell!} (1-x^2)^{m/2}
    {d^{\ell+m} \over dx^{\ell+m}} (x^2-1)^\ell \,,
 }
so that, for example, $Y_{00}=1/\sqrt{4\pi}$ and
 \eqn{YellmOne}{
  Y_{11} = -\sqrt{3 \over 8\pi} \sin\theta \, e^{i\phi} \qquad
  Y_{10} = \sqrt{3 \over \pi} \cos\theta \qquad
  Y_{1,-1} = \sqrt{3 \over 8\pi} \sin\theta \, e^{-i\phi} \,.
 }
Also, it is convenient to introduce functions
 \eqn{PhiNL}{
  \Phi_{n\ell}(\chi) = {d^{\ell+1} \cos n\chi \over
    d(\cos\chi)^{\ell+1}} \,,
 }
which are non-zero for $\ell < n$.

Spherical harmonics
 \eqn{Snlm}{
  \mathbb{S}(n\ell m) =
    {\sin^\ell \chi \over \sqrt{a_{n\ell m}}}
      \, \Phi_{n\ell}(\chi) Y_{\ell m}(\theta,\phi) \qquad
  a_{n\ell m} = {n\pi \over 2} {(\ell+n)! \over (n-\ell-1)!}
 }
obey \eno{DefineHSHscalar} with $k_S^2 = n^2-1$, where $|m| \leq \ell < n$.  The $\mathbb{S}(n\ell m)$ for fixed $n$ form $n^2$-dimensional representations of $SO(4) = SU(2)_L \times SU(2)_R$, namely spin-$(n-1)/2$ for both $SU(2)$ factors.  The $S_{n\ell m}$ are orthonormal:
 \eqn{SnlmNorm}{
  \int_{S^3} \sin^2 \chi \sin\theta \, d\chi \, d\theta \, d\phi
   \; \mathbb{S}(n\ell m) \, \mathbb{S}^*(n'\ell' m') =
    \delta_{nn'} \delta_{\ell\ell'} \delta_{mm'} \,.
 }
The first few $\mathbb{S}(n\ell m)$ are
 \eqn[c]{FirstFewScalar}{
  \mathbb{S}(100) = 1/\pi\sqrt{2} \qquad
  \mathbb{S}(200) = {\sqrt{2} \over \pi} \cos\chi  \cr
  \mathbb{S}(211) = -{1 \over \pi} \sin\chi \sin\theta \,
    e^{i\phi} \qquad
  \mathbb{S}(210) = {\sqrt{2} \over \pi} \sin\chi \cos\theta
    \qquad
  \mathbb{S}(2,1,-1) = {1 \over \pi} \sin\chi \sin\theta \,
    e^{-i\phi} \,.
 }

The group theory for harmonics on $S^3$ of conserved, symmetric, traceless tensors of rank $r$ is simple to state.  (Vector spherical harmonics are the special case $r=1$, and tensor spherical harmonics are the case $r=2$.)  Let $n_L$ denote the $n$-dimensional representation of $SU(2)_L$ (i.e.~spin $(n-1)/2$), and similarly for $SU(2)_R$, so that the $\mathbb{S}(n\ell m)$ for fixed $n$ fill out the representation $n_L \,\otimes\, n_R$.  Then, for fixed $n$, defined so that the eigenvalue of the laplacian is $-k_r^2 = -n^2+r+1$, with integer $n>r$, the representation content of the harmonics with that eigenvalue is
 \eqn{RepContent}{
  (n+r)_L \otimes (n-r)_R \;\oplus\; (n-r)_L \otimes (n+r)_R \,,
 }
for a total dimension of $2(n^2-r^2)$.  But there is a ${\bf Z}_2$ parity symmetry under which $SU(2)_L$ and $SU(2)_R$ are interchanged, and the representation \eno{RepContent} may be decomposed into symmetric and anti-symmetric product pieces, each with dimension $n^2-r^2$.

Explicit expressions for vector spherical harmonics were given in \cite{Tomita:1982ew} and are reproduced here.  For the even modes (corresponding to the symmetric part of the product \eno{RepContent}),
 \eqn[c]{VnlmEven}{
  \mathbb{V}^{\rm even}_j(n\ell m) = \begin{pmatrix} V_1^{n\ell} \\
       V_2^{n\ell} \partial_\theta \\ V_2^{n\ell} \partial_\phi
     \end{pmatrix} Y_{\ell m}(\theta,\phi)  \cr
  V_2^{n\ell} = {\sin^2 \chi \over \ell(\ell+1)} V_1^{n\ell} \qquad
  V_1^{n\ell} = \sqrt{\ell(\ell+1) \over n^2 a_{n\ell m}}
    \sin^{\ell-1} \chi \, \Phi_{n\ell}(\chi)
 }
where $a_{n\ell m}$ is defined as in \eno{Snlm} and $0 < \ell < n$.  The eigenvalue for these modes is $k_V^2 = n^2-2$.  The components of the vector are in the $\chi$, $\theta$, $\phi$ directions, and the index is down, meaning that $\mathbb{V}_i dy^i$ is a one-form.  Thus, for example,  $\mathbb{V}_i = {\tiny \begin{pmatrix} 1 \\ 0 \\ 0 \end{pmatrix}}$ means that $\mathbb{V}_i dy^i = d\chi$.  The $\mathbb{V}^{\rm even}_j(n\ell m)$ are orthonormal:
 \eqn{VnlmEvenNorm}{
  \int_{S^3} \sin^2 \chi \sin\theta \, d\chi \, d\theta \, d\phi
   \; \mathbb{V}^{{\rm even},j}(n\ell m) \,
    \mathbb{V}^{{\rm even},*}_j(n'\ell' m') =
    \delta_{nn'} \delta_{\ell\ell'} \delta_{mm'} \,,
 }
where the index $j$ on the first factor is raised using the standard metric on the unit $S^3$.  The first few $\mathbb{V}^{\rm even}_j(n\ell m)$ are
 \eqn[c]{FirstFewEvenVector}{
  \mathbb{V}^{\rm even}_j(211) =
   -{e^{i\phi} \over \pi\sqrt{2}} \begin{pmatrix}
     \sin\theta \\ \sin\chi \cos\chi \cos\theta \\
      i \sin\chi \cos\chi \sin\theta \end{pmatrix} \qquad
  \mathbb{V}^{\rm even}_j(210) = {1 \over \pi} \begin{pmatrix}
     \cos\theta \\ -\sin\chi \cos\chi \sin\theta \\ 0
    \end{pmatrix}  \cr
  \mathbb{V}^{\rm even}_j(2,1,-1) =
    {e^{-i\phi} \over \pi\sqrt{2}} \begin{pmatrix}
      \sin\theta \\ \sin\chi \cos\chi \cos\theta \\
       -i \sin\chi \cos\chi \sin\theta \end{pmatrix} \,.
 }
The odd vector modes are
 \eqn{VnlmOdd}{
  \mathbb{V}^{\rm odd}_j(n\ell m) =
    {\sin^{\ell+1} \chi \over \sqrt{\ell (\ell+1) a_{n\ell m}}} \,
    \Phi_{n\ell}(\chi) \, {1 \over \sin\theta} \begin{pmatrix}
       0 \\ -\partial_\phi \\ \sin^2 \theta \, \partial_\theta
     \end{pmatrix} Y_{\ell m}(\theta,\phi) \,,
 }
where we have altered the overall normalization from the one stated in \cite{Tomita:1982ew}, which appears to be slightly in error.  Again, $0 < \ell < n$, and the eigenvalue is $k_V^2 = n^2-2$.  The $\mathbb{V}^{\rm odd}_j(n\ell m)$ are orthonormal: a formula precisely analogous to \eno{VnlmEvenNorm} applies.  And they are orthogonal to the $\mathbb{V}^{\rm even}_j(n\ell m)$.  The first few $\mathbb{V}^{\rm odd}_j(n\ell m)$ are
 \eqn[c]{FirstFewOddVector}{
  \mathbb{V}^{\rm odd}_j(211) =
    {e^{i\phi} \over \pi\sqrt{2}} \begin{pmatrix}
       0 \\ i \sin^2\chi \\ -\sin^2\chi \sin\theta \cos\theta
     \end{pmatrix} \qquad
  \mathbb{V}^{\rm odd}_j(210) = -{1 \over \pi} \begin{pmatrix}
    0 \\ 0 \\ \sin^2\chi \sin^2\theta \end{pmatrix}  \cr
  \mathbb{V}^{\rm odd}_j(2,1,-1) =
    {e^{-i\phi} \over \pi\sqrt{2}} \begin{pmatrix}
       0 \\ i \sin^2\chi \\ \sin^2\chi \sin\theta \cos\theta
     \end{pmatrix} \,.
 }
It is easy to verify that there are $n^2-1$ even and $n^2-1$ odd vector harmonics at level $n \geq 2$, as the representation theory \eno{RepContent} demands.

Even tensor harmonics, again following \cite{Tomita:1982ew}, take the form
 \eqn{TnlmEven}{
  \mathbb{T}^{\rm even}_{ij}(n\ell m) =
   \begin{pmatrix} T_1^{n\ell} & T_2^{n\ell} \partial_\theta &
      T_2^{n\ell} \partial_\phi  \\
    T_2^{n\ell} \partial_\theta & T_3^{n\ell} \partial_\theta^2 +
     T_4^{n\ell} &
     T_3^{n\ell} (\partial_\theta -\cot\theta) \partial_\phi \\
    T_2^{n\ell} \partial_\phi &
     T_3^{n\ell} (\partial_\theta -\cot\theta) \partial_\phi &
     T_3^{n\ell} (\partial_\phi^2 + \sin\theta \cos\theta
       \partial_\theta) + T_4^{n\ell} \sin^2\theta
   \end{pmatrix} Y_{\ell m}(\theta,\phi)
 }
where
 \eqn{TcoefsEven}{
  T_1^{n\ell} &= \sqrt{(\ell-1) \ell (\ell+1)(\ell+2) \over
    2n^2 (n^2-1) a_{n\ell m}} \, \sin^{\ell-2} \chi \,
      \Phi_{n\ell}(\chi)  \cr
  T_2^{n\ell} &= {\sin^2\chi \over \ell(\ell+1)}
    (\partial_\chi + 3\cot\chi) T_1^{n\ell}  \cr
  T_3^{n\ell} &= {\sin^2\chi \over (\ell-1)(\ell+2)}
    \left[ 2 (\partial_\chi + 2\cot\chi) T_2 - T_1^{n\ell} \right]
      \cr
  T_4^{n\ell} &= {1 \over 2} \left[ \ell(\ell+1) T_3^{n\ell} -
    \sin^2\chi \, T_1^{n\ell} \right]
 }
and $n > \ell > 1$.  The eigenvalue for these modes is $k_T^2 = n^2-3$.  The $\mathbb{T}^{\rm even}_{ij}(n\ell m)$ are orthonormal:
 \eqn{TnlmEvenNorm}{
  \int_{S^3} \sin^2 \chi \sin\theta \, d\chi \, d\theta \, d\phi
   \; \mathbb{T}^{{\rm even},ij}(n\ell m) \,
    \mathbb{T}^{{\rm even},*}_{ij}(n'\ell' m') =
    \delta_{nn'} \delta_{\ell\ell'} \delta_{mm'} \,,
 }
The $n=3$ even tensor spherical harmonics, $\mathbb{T}^{\rm even}_{ij}(3,2,m)$, fill out a $5$-dimensional representation of $SO(4)$.  The explicit form of these tensors is too complicated to be worth displaying explicitly here.

Odd tensor spherical harmonics take the form
 \eqn{TnlmOdd}{
  &\mathbb{T}^{\rm odd}_{ij}(n\ell m) =  \cr
   &\begin{pmatrix} 0 & -T_5^{n\ell} \csc\theta \, \partial_\phi &
      T_5^{n\ell} \sin\theta \, \partial_\theta \\
    -T_5^{n\ell} \csc\theta \, \partial_\phi &
     -2 T_6^{n\ell} \csc\theta (\partial_\theta - \cot\theta)
       \partial_\phi &
     T_6^{n\ell} (\sin\theta \, \partial_\theta^2 -
      \cos\theta \, \partial_\theta -
      \csc\theta \, \partial_\phi^2) \\
    T_5^{n\ell} \sin\theta \, \partial_\theta &
     T_6^{n\ell} (\sin\theta \, \partial_\theta^2 -
      \cos\theta \, \partial_\theta -
      \csc\theta \, \partial_\phi^2) &
     2 T_6^{n\ell} \sin\theta (\partial_\theta - \cot\theta)
      \partial_\phi
   \end{pmatrix}  \cr
   &\qquad\quad{} \times Y_{\ell m}(\theta,\phi)
 }
where
 \eqn{TcoefsOdd}{
  T_5^{n\ell} &= \sqrt{(\ell-1)(\ell+2) \over
    2\ell (\ell+1) (n^2-1) a_{n\ell m}} \, \sin^\ell \chi \,
      \Phi_{n\ell}(\chi)  \cr
  T_6^{n\ell} &= {1 \over (\ell-1)(\ell+1)} \partial_\chi
    (\sin^2 \chi \, T_5^{n\ell}) \,.
 }
Again $n > \ell > 1$, and $k_T^2 = n^2-3$.  We have altered the overall normalization on $\mathbb{T}^{\rm odd}_{ij}(n\ell m)$ from the expressions given in \cite{Tomita:1982ew}, which appear to be slightly in error.  The $\mathbb{T}^{\rm odd}_{ij}(n\ell m)$ are orthonormal, and they are orthogonal to the $\mathbb{T}^{\rm even}_{ij}(n\ell m)$.  The expressions for even the low-lying $\mathbb{T}^{\rm odd}_{ij}(n\ell m)$ are too complicated to reproduce here.  As with the vector modes, it is easy to verify that there are $n^2-4$ even and $n^2-4$ odd tensor harmonics at level $n \geq 3$, as the representation theory \eno{RepContent} demands.

\clearpage

\bibliographystyle{ssg}
\bibliography{bh}

\end{document}